\documentclass[11pt, a4paper]{article}

\usepackage[height=8.85in,width=6.45in]{geometry}

\usepackage{graphicx,rotating}     
\usepackage[bookmarksopen,colorlinks=true,linkcolor=light_blue,
citecolor=light_pink,urlcolor=dark_red,linktocpage=false]{hyperref}

\usepackage{amsfonts}
\usepackage{amsmath}
\usepackage{amssymb}
\usepackage{slashed}
\usepackage{braket}
\usepackage{subfigure}
\usepackage{bm,here}
\usepackage{cite}
\usepackage{comment}
\usepackage{cases}
\usepackage{colortbl}
\usepackage{xcolor}
\usepackage[utf8]{inputenc}
\usepackage[mathscr]{eucal}
\usepackage[footnotesize]{caption}
\usepackage{lipsum}
\usepackage{subfigure}

\usepackage[utopia]{mathdesign}

\usepackage{multicol}
\definecolor{dark_blue}{rgb}{0.0, 0., 0.6}
\definecolor{dark_red}{rgb}{0.7, 0., 0.}
\definecolor{dark_green}{rgb}{0., 0.45, 0.3}
\definecolor{light_pink}{rgb}{1,0.4,0.4}
\definecolor{light_blue}{rgb}{0.284602,0.317763,0.963947}
\definecolor{red}{rgb}{1,0,0}
\definecolor{blue}{rgb}{0,0,1}
\definecolor{orange}{rgb}{1,0.5,0}

\newcommand{\nn}{\nonumber \\}
\newcommand{\lmk}{\left(}  
\newcommand{\rmk}{\right)}
\newcommand{\lkk}{\left[}  
\newcommand{\rkk}{\right]}

\newcommand{\del}{\partial}

\newcommand{\bea}{\begin{array}}
\newcommand{\eea}{\end{array}}
\newcommand{\beq}{\begin{eqnarray}}
\newcommand{\eeq}{\end{eqnarray}}
\newcommand{\eq}[1]{Eq.~(\ref{#1})}

\newcommand{\cphi}{\varphi}

\newcommand{\dd}{\mathrm{d}}

\newcommand{\abs}[1]{\left\vert {#1} \right\vert}

\makeatletter
\@addtoreset{equation}{section}

\makeatother

\usepackage{framed}
\definecolor{shadecolor}{rgb}{0.95,0.95,0.95}

\definecolor{cred}{RGB}{180,50,40} 
\definecolor{purple}{RGB}{180,90,180} 
\definecolor{darkgreen}{RGB}{0, 100, 0}

\begin{document}

\hypersetup{pageanchor=false}
\begin{titlepage}

\begin{center}

\hfill DESY 18-124\\

\vskip 1.in

{\huge \bfseries 
Classical Nonrelativistic Effective Field Theory\\
and the Role of Gravitational Interactions\\
}

\vskip .5in

{\Large Joshua Eby$^{\spadesuit}$, Kyohei Mukaida$^{\diamondsuit}$, Masahiro Takimoto$^{\spadesuit}$, \\[.4em] L.~C.~R.~Wijewardhana$^{\clubsuit}$, Masaki Yamada$^{\heartsuit}$}

\vskip .3in
\begin{tabular}{ll}
$^{\spadesuit}$ &\!\! \emph{Department of Particle Physics and Astrophysics, }\\
&\!\! \emph{Weizmann Institute of Science, Rehovot 761001, Israel}\\[.3em]
$^{\diamondsuit}$&\!\! \emph{DESY, Notkestra{\ss}e 85, D-22607 Hamburg, Germany}\\[.3em]
$^{\clubsuit}$&\!\! \emph{Department of Physics, University of Cincinnati, Cincinnati, OH 45221, USA}\\[.3em]
$^{\heartsuit}$ &\!\! {\em Institute of Cosmology, Department of Physics and Astronomy,}\\
&\!{\em Tufts University, Medford, MA  02155, USA}
\end{tabular}

\end{center}
\vskip .6in

\begin{abstract}
\noindent
Coherent oscillation of axions or axion-like particles may give rise to long-lived clumps, called axion stars, because of the attractive gravitational force or its self-interaction. Such a kind of configuration has been extensively studied in the context of oscillons without the effect of gravity, and its stability can be understood by an approximate conservation of particle number in a non-relativistic effective field theory (EFT). We extend this analysis to the case with gravity to discuss the longevity of axion stars and clarify the EFT expansion scheme in terms of gradient energy and Newton's constant. Our EFT is useful to calculate the axion star configuration and its classical lifetime without any ad hoc assumption. In addition, we derive a simple stability condition against small perturbations. Finally, we discuss the consistency of other non-relativistic effective field theories proposed in the literature. 
\end{abstract}

\end{titlepage}

\tableofcontents
\thispagestyle{empty}
\renewcommand{\thepage}{\arabic{page}}
\renewcommand{\thefootnote}{$\natural$\arabic{footnote}}
\setcounter{footnote}{0}
\newpage
\hypersetup{pageanchor=true}

\section{Introduction}

Light scalar particles arise in numerous theories of physics beyond the Standard Model. A prime example is the axion, typically realized as a pseudo Nambu-Goldstone boson (pNGB) of a broken $U(1)$ symmetry at some high cosmological scale \cite{Peccei:1977hh, Peccei:1977ur, Weinberg:1977ma}. Such particles are also natural candidates for the particle nature of dark matter \cite{Preskill:1982cy, Sikivie:2006ni}. 

Axions are produced early in the universe, and naturally occupy states with very high occupation numbers, that coherently oscillate with dominant frequency $\omega \sim m_\phi$, where $m_\phi$ is the axion mass. Such states are well-described by a classical field. Owing to the smallness of $m_\phi$, higher-order contributions to the oscillation frequency $\omega_n \sim n\,m_\phi$ are extremely rapid and are typically neglected at leading order. There now exists a standard procedure for deriving the low energy limit of the Klein-Gordon equation for a real scalar field: one expands the relativistic field $\phi$ in a nonrelativistic wavefunction $\psi$ using the relation
\begin{equation} \label{NRlimit}
 \phi(t,{\bm x}) = \frac{1}{\sqrt{2m_\phi}}\left[e^{-i\,m_\phi\,t}\psi(t,{\bm x}) + e^{i\,m_\phi\,t}\psi^*(t,{\bm x})\right],
\end{equation}
and drops any term beyond leading order in rapidly oscillating factors $e^{\pm i\,m_\phi\,t}$. The resulting equation of motion, generically a nonlinear Schr\"odinger equation, is both classical and non-relativistic.

Localized quasi-stable solutions to the classical, non-relativistic equations of motion are known as \emph{oscillons} \cite{Gleiser:2009ys,Hertzberg:2010yz,Mukaida:2016hwd} or \emph{boson stars}.\footnote{In this work, we will use these terms interchangeably, though the usual terminology is that boson stars are coupled to gravity while oscillons are not. Sometimes the term \emph{oscillaton} is used to refer to an oscillon coupled to gravity \cite{Seidel:1991zh,Seidel:1993zk}.} Such objects can be supported by a balance of attractive and repulsive forces, sometimes including gravity. The original solutions for gravitationally bound (but otherwise non-interacting) boson stars were found by \cite{Kaup:1968zz,Ruffini:1969qy}, though self-interactions have been included in recent years either in generic $\phi^4$ scalar theory \cite{Colpi:1986ye,Chavanis:2011zi,Chavanis:2011zm}, and also in the specific case of the axion potential \cite{Barranco:2010ib,Eby:2014fya}. In the latter case then configurations are referred to as axion condensates, or more often, \emph{axion stars}.

There are certain applications in which the non-relativistic limit may be insufficient. Over the last few years, a number of procedures have appeared for organizing corrections to the non-relativistic limit, which can generically be referred to as non-relativistic effective field theories (NREFTs). One such method was presented by some of the present authors (hereafter MTY) \cite{Mukaida:2016hwd}, in which the scalar field was decomposed into non-relativistic and rapidly oscillating parts, $\phi = \phi_\text{NR} + \delta\phi$. MTY presented a scheme for integrating out $\delta\phi$ perturbatively in the self-interaction coupling. This gave rise to corrections to the self-interaction couplings, as well as new terms in the equations of motion which were higher-order in spatial gradients as well as time derivatives. The method of MTY thereby accounted for all relevant relativistic corrections.
Some relativistic corrections give imaginary terms in the EFT, which leads to the decay of oscillon states \cite{Hertzberg:2010yz,Eby:2015hyx,Braaten:2016dlp,Visinelli:2017ooc}. The lifetime of an oscillon can be estimated from the EFT and the result is consistent with the classical numerical simulation. 

A different method for constructing the NREFT of a real scalar field was performed by Braaten, Mohapatra, and Zhang (hereafter BMZ) \cite{Braaten:2016kzc}, who matched relativistic scattering amplitudes at high energies to effective operators in the low energy theory, giving rise to modified self-interaction couplings. The original work of BMZ did not take into account higher-order gradients and time derivatives, which give important contributions at the same order in the EFT expansion as the self-interaction corrections they calculated. In a recent update \cite{Braaten:2018lmj}, they have presented a more complete NREFT which includes these corrections as well.

Very recently, yet another method was presented by Namjoo, Guth, and Kaiser (hereafter NGK) \cite{Namjoo:2017nia}. In this work, NGK defined the relativistic field using a nonlocal operator which generalized Eq. \eqref{NRlimit}, then expanded the wavefunction $\psi$ and its conjugate $\psi^*$ as a tower of oscillating modes with frequencies $\pm n\,m_\phi$. The equation of motion for the lowest order mode, taken to have frequency $+m_\phi$, was determined by integrating out the other modes.\footnote{As this paper was being finalized, an update to \cite{Namjoo:2017nia} appeared which clarified the expansion scheme in their method by separating self-interaction, spatial derivative, and time derivative corrections.} It has been shown that, under appropriate field redefinitions, the methods of MTY, BMZ, and NGK give rise to matching elements in the low-energy S-matrix at $\psi^6$ order \cite{Namjoo:2017nia} and in the T-matrix at $\psi^8$ order \cite{Braaten:2018lmj}.

A typical boson star is dilute and weakly bound, and as a result, relativistic corrections are extremely small. A theoretically ideal application in which relativistic corrections become important is in a \emph{dense axion star}; in this configuration, the binding energy of the axions is large and the radius of the star is extremely small (close to the Compton wavelength of individual axions) \cite{Braaten:2015eeu}. This is a clear scenario in which the approximation of neglecting rapidly oscillating terms breaks down badly, and one must instead integrate out these modes. In \cite{Eby:2017teq}, the leading nontrivial corrections to the wavefunctions of dense axion stars were calculated by generalizing the field operator for the scalar to include these fast-oscillating modes. The resulting self-interaction potential also agreed well with the effective interactions calculated in \cite{Mukaida:2016hwd} at $\psi^6$ order. Importantly, in the dense axion star, gravity does not play an important role, as the self-interactions terms dominate \cite{Eby:2016cnq,Visinelli:2017ooc}, and so the application of a non-gravitational NREFT was appropriate. This is not the case in a dilute axion star, where gravity contributes at leading order. To analyze such states in an NREFT, gravity must be taken into account.

Thus, today there exist a plethora of methods of calculation in the NREFT of real scalar fields. It is relevant to understand the consistency of the results between methods, and also to point out any relevant advantages of one over another.
In this note, we will clarify the general MTY procedure for calculating corrections in NREFT \cite{Mukaida:2016hwd}. 
We will argue for a number of advantages to this method as compared to that of NGK or BMZ. One such advantage of the MTY method is that, as we will describe in this note, it is straightforward to include the effect of gravity, though in all cases it gives rise to certain theoretical complications. A thorough treatment of the gravitational interaction is especially important for understanding relativistic corrections to many classes of boson stars, where gravity contributes importantly at leading order.

As the axion field is a Hermitian (real) scalar, there is no symmetry which protects axion stars against number changing interactions. Nevertheless, there is an approximately conserved particle number that renders the configuration extremely long-lived \cite{Eby:2015hyx,Mukaida:2016hwd}. The NREFT presented here renders the calculation of the classical decay rate for emission of high-energy particles extremely straightforward. As a concrete example, we apply this method to estimate the lifetime of dense axion stars.

One way to establish the consistency of an NREFT is to compare a numerical oscillon solution and prediction by EFT; this was established for the MTY method in~\cite{Mukaida:2016hwd}. Another way to check the consistency is to compare an exact solution and prediction by EFT. There are a few situations where an exact solution is known; one example is spatially homogeneous $\phi^4$ theory, where the exact solution is given by an elliptic function. We will compare this exact solution with the result from the MTY, BMZ, and NGK effective field theory calculations.

The paper is organized as follows. In Section \ref{sec:EFT}, we present the calculation of relativistic corrections in NREFT as an expansion in the small parameters of the theory, building on the MTY formalism of \cite{Mukaida:2016hwd}. In Section \ref{sec:EFT_Grav}, we describe how to couple this theory to gravity, and compare corrections coming from gravity to the other expansion parameters. We explain how to calculate the imaginary part of the Lagrangian, which gives rise to decay processes even in the classical regime, in Section \ref{sec:decay}. Finally, we compare various NREFT methods in Section \ref{sec:comparison}, and we conclude in Section \ref{sec:Conclusions}.

We use natural units throughout, where $\hbar = c = 1$.

\section{NR EFT without gravity}
\label{sec:EFT}

In this section, we explain the method of calculation in our NREFT, building on our original work \cite{Mukaida:2016hwd}. 
The advantage of this method in the study of oscillons is sixfold: 
the calculation is easy and straightforward by using Feynman diagrams; 
the reason for (approximate) stability is clear; 
it is easy to calculate the background configuration; 
it is straightforward to include the effect of gravity; 
the lifetime can be calculated from the imaginary part of the Lagrangian
; and the higher order corrections in the small parameter expansion become small. Each of these points will be illustrated in what follows.

We start from the following Lagrangian for a relativistic quantum field theory of a real scalar field $\phi$: 
\begin{align}
         \mathcal{L}=\frac{1}{2}(\partial\phi)^2-\frac{1}{2}m^2_\phi\phi^2-V_{\rm int}(\phi^2),
         \label{setup}
\end{align}
where we assume a $\mathbb{Z}_2$ symmetry for representative simplicity. 
The case without $\mathbb{Z}_2$ symmetry has been discussed in Ref.~\cite{Mukaida:2016hwd}, 
though in that work we did not include the effect of gravity. 
In this section, we refine the method and clarify the expansion scheme of our NREFT. 
In Section \ref{sec:EFT_Grav}, we will explain how to include the effect of gravity. 

\subsection{Expansion scheme}
We first decompose $\phi$ into slowly and fast oscillating parts:
\begin{align}
         \phi(t, \bm{x})= 
		\left[ e^{- i \omega t} \psi_1 (t, \bm{x}) + \text{h.c.} \right] + \delta \phi, 
\label{decomposition1}
\end{align}
where the frequency $\omega$ ($\simeq m_\phi$) will be determined later.
We sometimes expand the fast oscillating parts as
\begin{align}
	\delta\phi \equiv \sum_{n \geq 2} e^{- i n \omega t} \psi_n
	+ \text{h.c}.,
	\label{mode_rel}
\end{align}
where each $\psi_n$ is assumed to be a slowly varying field.%
\footnote{
We use a different unit of frequency for Fourier modes and a different normalization for $\psi_n$ from those used in our previous paper~\cite{Mukaida:2016hwd} 
so that the resulting NREFT is simpler. 
} 
Note that this is just a Fourier decomposition 
and each $\psi_n$ is the mode that has a positive frequency. 
The goal is to derive the EFT by assuming $\psi_1$ is the dominant mode (in the non-relativistic regime) and integrating out the other highly oscillating modes $\psi_{n>1}$.

It is instructive to mention the implicit limitation of the mode expansion in Eq. \eqref{mode_rel} here.
This expansion of the relativistic component $\delta \phi$ is useful when we compute the \emph{classical} NREFT, defined by tree-level diagrams. Note however that this is not true if we would like to consider quantum effects because the energy of $\delta \phi$ must be continuous in loop diagrams. For this purpose we have to stop the decomposition of the scalar field at $\phi = (e^{-i \omega t}\psi_1 + \text{h.c.}) + \delta \phi$. See Ref.~\cite{Mukaida:2016hwd} for the definition of $\delta \phi$ in this case. The case of quantum decay for axion stars was also considered previously in \cite{Braaten:2016dlp}.

Let us also clarify the role of $\omega$ here.
In Sec.~\ref{stable} we discuss how to get a stationary configuration of $\psi_1$. There we will take $\omega$ so that it factors out the time dependence of $\psi_1$.
Nevertheless we sometimes keep both $\partial_t \psi_1$ and $\omega$, since one may also take $\omega \to m_\phi$ and keep $\partial_t\psi_1$ instead. The latter limit is useful for computing scatterings of \textit{free} $\psi_1$ particles as done in Ref.~\cite{Braaten:2018lmj}. We emphasize that both are equivalent after solving the equation of motion for $\psi_1$. We clarify that we do not need to include a time-dependent term by hand in our EFT, as suggested in Ref.~\cite{Braaten:2018lmj}. Such terms are already taken into account in Ref.~\cite{Mukaida:2016hwd}. We demonstrate the equivalence of each EFT in Sec.~\ref{sec:comparison} by comparing with an exact solution in the limit of $\nabla \psi_1 \to 0$.
Also, in Sec.~\ref{sec:timederivative}, we explicitly show that our EFT contains the time derivative term pointed out in Ref.~\cite{Braaten:2018lmj}, if one takes $\omega \to m_\phi$ and keeps $\partial_t \psi_1$.

The EFT method is useful for the study of a clump of oscillating scalar field, like an oscillon or axion star. 
In the non-relativistic regime, there are three possible small quantities in the equation of motion:
\beq
         &&\delta_x \equiv \frac{\abs{{\bm\nabla}^2\psi_1}}{\abs{m_\phi^2 \psi_1}} \ll 1,\label{small del_x} \\
         &&\delta_V \equiv \frac{\abs{V_{\rm int}(\abs{\psi_1}^2)}}{m_\phi ^2\abs{\psi_1}^2} \ll 1, \\
         &&\delta_t \equiv \abs{\frac{m_\phi - \omega - i\frac{\del_t\psi_1}{\psi_1}}{m_\phi}} \ll 1. 
\eeq
For each of them, we can assign pseudo parameters $\epsilon_{x,V,t}=1$
like
\begin{align}
         \abs{{\bm\nabla}^2 \psi_1}&\rightarrow \epsilon_x \abs{{\bm\nabla}^2 \psi_1},\\
         V_{\rm int}&\rightarrow \epsilon_V V_{\rm int}, \\
	 (m_\phi - \omega) &\rightarrow \epsilon_t (m_\phi - \omega). 
\end{align}
Then, we can systematically estimate relativistic modes by using an expansion in
 $\epsilon_*$. As a result, a term which is proportional to $(\epsilon_x)^a\,(\epsilon_t)^b\,(\epsilon_V)^c$ in the resulting effective Lagrangian will have relative size $(\delta_x)^a\,(\delta_t)^b\,(\delta_V)^c$, where $a,b,c\in\mathbb{Z}$ are exponents of the expansion parameters. This fact supports the validity of an expansion in $\epsilon_*$, given that $\delta_*$ are small.
 
There are certain oscillon solutions for which we find the following relation,
\begin{align}
\label{rere}
         m_\phi (m_\phi - \omega) \sim \frac{|{\bm\nabla}^2\psi_1|}{\abs{\psi_1}}
         			\sim \frac{V_{\rm int}(\abs{\psi_1}^2)}{\abs{\psi_1}^2}, 
\end{align}
which implies $\delta_t  \sim \delta_x \sim \delta_V$. 
In such a case, the resulting effective Lagrangian can be
decomposed into powers of a single expansion parameter ($\epsilon_V$ for example). 
In general, the EFT will contain higher derivative terms with respect to time, 
but we can use the equation of motion to remove them (see Refs.~\cite{Arzt:1993gz, GrosseKnetter:1993td}).

Let us illustrate how to obtain the NREFT here.
Our strategy to get the NREFT is straightforward: keep the NR mode, $\psi_1$, and integrate out the fast oscillating parts, $\delta \phi$. The EFT action can be decomposed into kinetic, mass, and interaction parts
\begin{equation}
 S_{NR} = S_{\text{kin}} + S_{\text{mass}} + S_{\text{int}}.
\end{equation}
First of all, substituting this decomposition into Eq.~\eqref{setup},
one may readily find that the kinetic term plus the mass term become: 
\begin{align}
	S_\text{kin} + S_\text{mass} 
	&= \int_{x} \left[ \abs{\partial_t \psi_1 - i \omega \psi_1}^2
	-\abs{\bm{\nabla} \psi_1} ^2 - m_\phi^2 \abs{\psi_1}^2
	- \frac{1}{2} \delta \phi \left( \Box + m_\phi^2 \right) \delta \phi
	\right]\\ 
	&= \int_{x}
	\sum_{n \geq 1} \psi_n^{\dag} \left(2 i n \omega \partial_t - m_\phi^2 + n^2 \omega^2  + \bm{\nabla}^2 - \partial_t^2 \right) \psi_n, \label{Skin+mass}
\end{align}
where $\int_x = \int d^4 x$ is the spacetime volume integration. 
Note here that there are no cross terms among different $n$ of $\psi_n$. 
This is because the $\psi_n$s are slowly oscillating by assumption, meaning that their time dependence is weak compared to the exponential prefactor. To make this point clear, let us consider the cross term of $\psi_{n'}^\dag$ and $\psi_n$ for $n \neq n'$. There, we expect terms proportional to $e^{i(n'-n) \omega t} \psi_{n'}^\dag \psi_n$. If $\psi_n$ and $\psi_{n'}$ are time independent, then the integral of such a term over time is proportional to a delta function $\delta\left((n'-n)\omega\right)$, which is nonzero only if $n'=n$. A weak dependence on time effectively shifts this delta function, but in such a way that its argument is never zero, and so such terms integrate to zero in this case as well. Hence, all the terms must appear in a pair of $\psi_n^\dag$ and $\psi_n$.

Then we move on to the interaction term.
To make our discussion concrete, let us consider the following potential as an example:
\begin{align}
	V_\text{int} = \frac{\lambda_4}{4!} \phi^4.
	\label{V4}
\end{align}
The fact that $\psi_n$s 
are slowly varying compared to
$e^{- i \omega t}$  is also essential for rewriting the interaction term by means of our decomposition. Taking this into account, one can decompose the interaction term into two parts:
\begin{align}
	S_\text{int} &= - \int_x 
		\frac{\lambda_4 \abs{\psi_1}^4}{4}
		+ \delta S [\psi_1, \delta \phi],
\end{align}
with 
\begin{align}
	\delta S \equiv &
	-\int_x \Bigg[ \frac{\lambda_4}{3 !} \psi_1^3 e^{- 3 i \omega t} \delta \phi
	+ \frac{3 \lambda_4}{2 \cdot 3 !} \left( \psi_1^2 e^{ - 2 i \omega t} \delta \phi^2 
	+ \abs{\psi_1}^2 \delta \phi^2 \right)
	+ \frac{\lambda_4}{3 !} \psi_1 e^{- i \omega t} \delta \phi^3  \Bigg] + \text{h.c.} \label{nr_rela}\\
	&- \int_x\frac{\lambda_4}{4 !} \delta \phi^4 \label{rela_self}  \\[.5em]
	 =&-\int_x \Bigg[ \frac{\lambda_4}{3 !} \psi_1^3 \psi_{-3}
	 + \frac{3 \lambda_4}{3 !}
	 \left( \psi_1^2 \sum_{j \geq 4} \psi_{-j} \psi_{j-2} 
	 	+\abs{\psi_1}^2 \sum_{j \geq 2} \abs{\psi_j}^2
	 \right)
	 	+  \frac{\lambda_4}{3 !}\psi_1 \!\!\!\!\!\!\!\!\!\!\sum\displaylimits_{\substack{i,j,k\neq \pm1 \\ i + j + k = -1}} 
		\!\!\!\!\!\!\!\! \psi_{i} \psi_{j} \psi_{k} \Bigg] + \text{h.c.} \\
		&- \int_x \frac{\lambda_4}{4 !} \!\!\!\!\sum\displaylimits_{\substack{i,j,k,l \neq \pm1 \\ i + j + k + l = 0}} 
		\!\!\!\!\!\!\!\! \psi_{i} \psi_{j} \psi_{k} \psi_{l}. 
\end{align}
Here we have adopted the following notation $\psi_{-n} = \psi_{n}^\dag$. The second term, $\delta S$, represents the interaction between the NR mode $\psi_1$ and the relativistic modes contained in $\delta \phi$. In the second equality, we have used the mode expansion of $\delta \phi$ given in Eq.~\eqref{mode_rel} which is useful to derive the NREFT in the classical case. Terms with uncompensated $e^{\pm i\,n\,\omega\,t}$ prefactors, e.g. $e^{-4\,i\,\omega\,t}\psi_1^4$, all integrate to zero, because of the slow time dependence of $\psi_n$ (see the discussion surrounding Eq. \eqref{Skin+mass}).

\begin{figure}
	\centering
	\includegraphics[width=.30\textwidth]{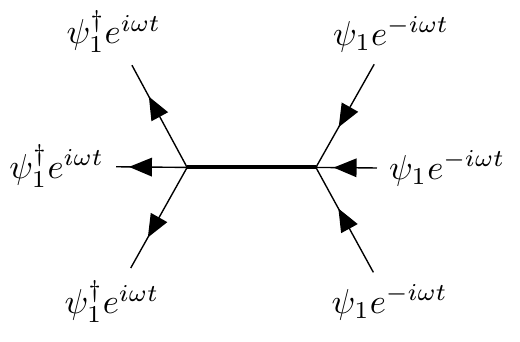}
	\caption{The leading order diagram at $\epsilon_V^2$ for the $\lambda_4 \phi^4$ interaction is shown. A thick line represents the relativistic mode, whose off-shell propagator is the inverse of the operator in parentheses in Eq. \eqref{Skin+mass}.}
	\label{fig:3to3}
\end{figure}

Now we are in a position to integrate out the relativistic mode $\delta \phi$. The only diagram that contributes up to $\epsilon_V^2$ is depicted in Fig.~\ref{fig:3to3}, namely a $3$-to-$3$ process mediated by one relativistic particle:
\begin{align}
	\delta S &\supset - \int_x \frac{\lambda_4^2}{(3 !)^2} \psi_1^\dag{}^3  \frac{1}{9 \omega^2  - m_\phi^2 + 6 i \omega \partial_t + \bm{\nabla}^2 - \partial_t^2 
	+ i \epsilon} \psi_1^3 \label{3to3} 
				+ \mathcal{O}(\epsilon_V^3,\epsilon_V^2 \epsilon_t, \epsilon_V^2\epsilon_x)\\
	& = - \int_x \frac{\lambda_4^2}{8 (3 !)^2} \frac{\abs{\psi_1}^6}{m_\phi^2} + \left(\text{imaginary part}\right) 
				+ \mathcal{O} (\epsilon_V^3,\epsilon_V^2 \epsilon_t, \epsilon_V^2\epsilon_x).
\end{align}
Note that the propagator for the internal line is off-shell, and can be read off from the expression of Eq. (\ref{Skin+mass}). Also, note that there exists a cut contribution that represents the production of relativistic particles. We will come back to this point in Sec.~\ref{sec:decay} later, but for now just recognize that we have an imaginary part in our EFT.

To sum up, the resulting EFT for $\psi_1$ takes the form of 
\beq
 S_{\rm NR} = \int_x 
 \lkk \abs{\del_t \psi_1 - i\omega \psi_1}^2 - \abs{\bm{\nabla} \psi_1}^2 
 - m_\phi^2 \abs{\psi_1}^2 
 - V_{\rm eff} (\abs{\psi_1}, \omega) - i \Gamma [\psi_1],
 \rkk. 
\label{EFT1}
\eeq
where $\Gamma[\psi_1]$ represents the imaginary part of the action, which is responsible for the lifetime of clumps; this will be discussed in Sec.~\ref{sec:decay}. 
Note that each term of the resulting EFT must contain the same number of $\psi_1$ and $\psi_1^\dag$. This, again, owes to the slow oscillation of $\psi_1$ compared to $e^{i\,\omega\,t}$: a vertex like $\psi_1^{n'} (\psi{_1^\dag})^{n} e^{-i \omega (n' - n) t}$ integrates to zero unless $n=n'$.\footnote{
 As long as the decay rate is suppressed, the frequency is sharply peaked and we expect this separation works well. As we will see in Section \ref{sec:decay}, the decay rate is strongly suppressed when $\omega \simeq m_\phi$. On the other hand, when $\omega \ll m_\phi$, the modes are no longer widely separated. 
}
Consequently, the final result just depends on $\abs{\psi_1}$.

For example, 
if the original interaction term is given by \eq{V4}, 
the effective potential is 
\beq 
 \left. V_{\rm eff} (\abs{\psi_1},\omega)\right\vert_{\epsilon_V^2} = 
 \frac{\lambda_4}{(2!)^2} \abs{\psi_1}^4 
 + \frac{1}{8} \frac{\lambda_4^2}{(3!)^2m_\phi^2} \abs{\psi_1}^6, 
 \label{eff1}
\eeq
up to $\epsilon_V^2$. 
As there are no corrections of the orders of $\epsilon_V \epsilon_x$ or $\epsilon_V \epsilon_t$, the corrections involving $\epsilon_t$ or $\epsilon_x$ start from $\epsilon_V^2 \epsilon_t$ or $\epsilon_V^2 \epsilon_x$: 
\beq
 &&\left. V_{\rm eff} (\abs{\psi_1},\omega) \right\vert_{\epsilon_V^2 \epsilon_t, \epsilon_V^2 \epsilon_x} 
 = 
 \frac{1}{8} \frac{\lambda_4^2}{(3!)^2 m_\phi^2} \psi_1^{*3} D_3 \psi_1^3,  
\eeq
where we defined derivative operators as 
\beq
 D_n \equiv - \frac{\bm\nabla^2}{(n^2-1) m_\phi^2} + \frac{n^2}{(n^2-1)} \lmk \frac{m_\phi^2 - \omega^2}{m_\phi^2} \rmk 
 + \frac{- 2n \omega i \del_t + \del_t^2}{(n^2-1) m_\phi^2}. 
 \label{Dn}
\eeq
The corrections up to $\epsilon_{x,V,t}^3$ is calculated in appendix~\ref{app:phi6}. 

Before leaving this section, we would like to emphasize that the calculation of the effective potential in this method is extremely straightforward. One can derive the EFT corrections directly using Feynman diagrams, or by considering the $\psi_{n>1}$ as perturbations on the equation of motion for $\psi_1$; both formulations give identical results. We illustrate the calculation procedure more fully in appendix \ref{app:eps3}, where we compute the effective potential at $\mathcal{O}(\epsilon^3)$.

\subsection{Stable configuration}
\label{stable}

If we neglect the imaginary part of the action, 
there is a global U(1) symmetry 
and the particle number of the field $\psi_1$ is conserved. 
In Ref.~\cite{Coleman:1985ki}, Coleman showed that 
there is a stable configuration in such a complex scalar field theory with a U(1) symmetry 
under some conditions. 
The stable solution is known as a Q-ball. 
Therefore 
scalar clumps in a real scalar field theory, like oscillon and axion star, 
can be understood as a projection of a Q-ball 
by the decomposition of \eq{decomposition1}. That is, neglecting higher-order modes which give rise to decay, the NR oscillon field is
\begin{equation}
 \phi_{NR} = 2\,\text{Re}\left[e^{-i\,\omega\,t}\psi_1\right],
\end{equation}
which is the projection on the real axis of the U(1)-symmetric Q-ball \cite{Mukaida:2014oza}.

Suppose that the number of particles in a system is given by $Q$, defined as
\begin{equation} \label{Qdef}
 Q = i \int_x 
 \lkk
 \psi_1^\dag \lmk \del_t - i \omega \rmk \psi_1 
 - \psi_1 \lmk \del_t + i \omega \rmk \psi_1^\dag
 \rkk .
\end{equation}
The most stable and energetically favourable configuration of the scalar field $\psi_1$ 
can be calculated by minimizing the energy with its number fixed. 
Hence we should minimize 
\beq
 \tilde{S}_\text{NR} = 
 &&\int_x 
 \lkk \abs{\del_t \psi_1 - i\omega \psi_1}^2 
 + \abs{\bm{\nabla} \psi_1}^2 
 + m_\phi^2 \abs{\psi_1}^2 
 + V_{\rm eff} (\abs{\psi_1}, \omega) 
 \rkk 
 \nn
 &&~~~~~~~~ + \omega' \lmk Q - i \int_x 
 \lkk
 \psi_1^\dag \lmk \del_t - i \omega \rmk \psi_1 
 - \psi_1 \lmk \del_t + i \omega \rmk \psi_1^\dag
 \rkk 
 \rmk, 
\eeq
where $\omega'$ is a Lagrange multiplier. 
This can be rewritten as 
\beq
 \tilde{S}_\text{NR} =
  \int_x 
 \lkk \abs{\del_t \psi_1}^2 
 + \abs{\bm{\nabla} \psi_1}^2 
 + (m_\phi^2 - \omega^2) \abs{\psi_1}^2 + V_{\rm eff} (\abs{\psi_1}, \omega) 
 \rkk 
 + \omega Q, 
\eeq
where we set $\omega = \omega'$. 
The first term is positive definite, 
so that it is minimized when $\del_t \psi_1(x,t) =0$. 
In other words, in the stable configuration we define $\omega$ so that $\del_t \psi_1 (t,{\bf x}) = 0$. 

Assuming spherical symmetry, 
we can determine the spatial dependence of $\psi_1$ by the following equation of motion: 
\beq
 \frac{\del^2 \psi_1}{\del r^2} + \frac{2}{r} \frac{\del \psi_1}{\del r} 
 - (m_\phi^2 - \omega^2) \psi_1 
 - \frac{\del V_{\rm eff}}{\del \psi_1^\dagger} = 0, 
 \label{EoM}
\eeq
where $r = |{\bm x}|$ is the radial coordinate. 
The boundary condition is $\psi_1 = 0$ for $r \to \infty$ and $\del_r \psi_1 = 0$ at $r = 0$. 
The condition for a solution to exist is as follows~\cite{Coleman:1985ki}: 
\beq
 - \frac{\del^2 V_{\rm eff}}{(\del \psi_1)(\del \psi_1^*)} (0)
 < m_\phi^2 - \omega^2 
 < - {\rm Min}_{\psi_1} \lkk \frac{V_{\rm eff}(\abs{\psi_1},\omega)}{\abs{\psi_1}^2} \rkk, 
\label{conditions}
\eeq
where ${\rm Min}_{\psi_1}[V_{\rm eff}/\abs{\psi_1}^2]$ represents the minimum of $V_{\rm eff}/\abs{\psi_1}^2$ as a function of $\abs{\psi_1}$. 
The solution of Eq. \eqref{EoM} is referred to as a Q-ball in the literature. 
 
Given $\del_t \psi_1 = 0$, the frequency $\omega$ and the particle number $Q$ are related by
\beq
 Q = \int_{\bm x} 2 \omega \abs{\psi_1}^2. 
 \label{Q}
\eeq
We can easily show that (see e.g. \cite{Gulamov:2013cra})
\beq
 \frac{\del E}{\del Q} = \omega, 
 \label{dEdQ}
\eeq
which means that the chemical potential inside of the Q-ball is equal to $\omega$. 
Therefore, 
if $\omega < m_\phi$, 
the Q-ball is stable compared with the free particle state.
In addition,
the stability condition against small perturbations is simply given by \cite{Friedberg:1976me,Lee:1991ax,Tsumagari:2009zp}
\beq
\frac{\omega}{Q}\frac{\dd Q}{\dd \omega} < 0.
\eeq
We derive this condition, even when gravitational interactions are included, in appendix~\ref{app:stab}.

Remembering \eq{decomposition1}, 
we can understand the oscillon configuration as a projection of a Q-ball, 
whose stability is guaranteed by a particle-number conservation. 
Note that we can determine the configuration of clumps 
from \eq{EoM}, which can be solved numerically 
by using the shooting method. 
The condition for the existence of clumps is also clear through \eq{conditions}. 
These are some of the advantages of our EFT compared with others used in the literature in the context of bound states of scalars. 

\section{NREFT including the effect of gravity} \label{sec:EFT_Grav}

\subsection{Effect of gravity on the scalar field}

Here we discuss the inclusion of gravity on the oscillon. The starting point is to promote the spacetime derivatives in the equation of motion to covariant derivatives,
\begin{equation}
 \Box \phi - m_\phi^2\,\phi - \frac{\partial V_\text{int}}{\partial\phi} = 0,
\end{equation}
where $\Box\phi = g^{\mu\nu}\partial_\mu \partial_\nu \phi - g^{\mu\nu} \Gamma^{\sigma}_{\mu\nu} \partial_\sigma \phi$ is the covariant d'Alembertian operating on $\phi$. Since we are interested in a spherically symmetric configuration, the metric around the oscillon can be written as
\begin{equation}
 \dd s^2 = A(t,r)\,\dd t^2 - B(t,r)\,\dd r^2 - r^2\left(\dd\theta^2 + \sin^2\theta\,\dd\varphi^2\right),
\end{equation}
where $A(t,r)$ and $B(t,r)$ are functions determined by the Einstein equation. 

The energy-momentum tensor of the (relativistic) scalar field is given by 
\beq
 T_{\mu \nu} = \del_\mu \phi \del_\nu \phi - \frac12 g_{\mu \nu} g^{\rho \sigma} 
 \del_\rho \phi \del_\sigma \phi - g_{\mu \nu} V_\text{int}(\phi). 
\eeq
The action of the scalar field is then given by
\beq
 S^{\rm (grav)} &&= \int_x \sqrt{-g} \lmk \frac12 g^{\mu \nu} \del_\mu \phi \del_\nu \phi 
 - \frac12 m_\phi^2 \phi^2 - V(\phi) \rmk
 \\
 &&= \int \dd t \dd r  \,4 \pi r^2 \sqrt{A\,B} \lmk \frac{1}{2 A} ( \del_t \phi)^2 
 - \frac{1}{2 B} \left(\frac{\del \phi}{\del r}\right)^2 - \frac12 m_\phi^2 \phi^2 - V_\text{int}(\phi) \rmk. 
\eeq
The EFT action for $\psi_1$ takes the form of 
\beq
 S_{\rm NR}^{(\rm grav)} = \int \dd t \dd r  \, 4 \pi r^2 \sqrt{A\,B} 
 \lkk \frac{1}{A} \abs{\del_t \psi_1 - i\omega \psi_1}^2 -  \frac{1}{B} \abs{\frac{\del \psi_1}{\del r}}^2 
 - m_\phi^2 \abs{\psi_1}^2 
 - V_{\rm eff} (\abs{\psi_1}, \omega) - i \Gamma [\psi_1]
 \rkk, 
 \nonumber\\
\label{EFT-gravity}
\eeq
where $V_{\rm eff}$ should be calculated as in the previous section. The gravitational corrections come through the factors of $A$ and $B$ in Eq. \eqref{EFT-gravity}.

We can write the energy-momentum tensor of the scalar field 
in the form of 
\beq
 T^\mu_\nu = {\rm diag} \lmk \rho, - p_r, - p_\perp, - p_\perp \rmk, 
 \label{EMtensor}
\eeq
where $\rho$ is the energy density, $p_r$ is the radial pressure, and $p_\perp$ is the tangential pressure~\cite{Ruffini:1969qy}, which in general are understood to be functions of the radial coordinate $r$ as well as time $t$. 
The $tt$- and $rr$-components of Einstein's equations lead to \cite{Weinberg:1972kfs}
\beq
 &&A(t,r) = \exp \lkk - 2 G \int_r^\infty \frac{\dd r'}{r'^2} \lmk M(t,r') + 4 \pi p_r r'^3 \rmk 
 \lmk 1 - \frac{2 G M(t,r')}{r'} \rmk^{-1} \rkk, 
 \label{A}
 \\
 &&B(t,r) = \lmk 1 - \frac{2 G M(t,r)}{r} \rmk^{-1}, 
 \label{B}
 \\
 &&M(t,r) \equiv \int_0^r d r' 4 \pi r'^2 \rho. \label{M}
\eeq
where $G = M_P{}^{-2}$ is the Newtonian gravitational constant and $M_P$ is the Planck mass.
The $\theta\theta$- and $\cphi\cphi$-components of Einstein's equations are automatically satisfied 
when the above equations and the equation of motion of the scalar field are satisfied.

\subsection{NR EFT with gravity corrections}

Now we assume that the effect of gravity is small and provide an expansion scheme. At leading order, 
we write $A(r) = 1 - 2\Psi(r)$ and $B(r) = 1 + 2 \Phi(r)$, and define the small parameter
\begin{equation}
 \delta_g \equiv \Psi \ll 1.
\end{equation}
We define the expansion in terms of $\Psi$ because, as we will see below, at leading order $\Phi$ decouples from the equations and only $\Psi$ is relevant. At higher order, we need to include $\Psi$ and $\Phi$. The parameter controlling the size of gravity corrections is Newton's gravitational constant, so we introduce a pseudo parameter $\epsilon_g=1$ via
\begin{equation}
 G  \rightarrow \epsilon_g\, G,
\end{equation}
as we did with the other corrections. 

When we define $\rho$ and $p_r$ by \eq{EMtensor}, 
they are determined by the scalar field as 
\begin{align}
 &\rho = \omega^2 \abs{\psi_1}^2 + \abs{\frac{\del \psi_1}{\del r}}^2 + m_\phi^2 \abs{\psi_1}^2 + V_{\rm eff}  \nonumber \\
 &~~~~ = 2 m_\phi^2 \abs{\psi_1}^2 \times \lkk 1 + {\cal O}(\epsilon_{x,V,t}) \rkk, 
 \\
 &p_r = \omega^2 \abs{\psi_1}^2 + \abs{\frac{\del \psi_1}{\del r}}^2 - m_\phi^2 \abs{\psi_1}^2 - V_{\rm eff} \nonumber \\
 &~~~~ = m_\phi^2 \abs{\psi_1}^2 \times  {\cal O}(\epsilon_{x, V, t}).
\end{align}
From Eqs.~(\ref{A}) and (\ref{B}), the gravitational potentials $\Phi$ and $\Psi$ are determined as 
\beq
 &&\Psi(t,r) =  G \int_r^\infty \frac{\dd r'}{r'^2} \lmk M(t,r') + 4 \pi p_r r'^3 \rmk, 
 \label{Psi}
 \\
 &&\Phi(t,r) = \frac{G M(t,r)}{r}, 
 \label{Phi}
\eeq
at the leading order in $\epsilon_{g}$.%
\footnote{
The tangential pressure $p_\perp$ is relevant in the $\theta\theta$-component of Einstein equation, but it can be obtained from the combination of Eqs.~(\ref{Psi}), (\ref{Phi}), and equation of motion. So we do not need to write down that equation~\cite{Gleiser:1988rq}. 
}
At the leading order in $\epsilon_{g,x,t}$,
the effective action and Lagrangian is simply given by
\begin{align}
S&=\int \dd t \dd r (4\pi r^2) \mathcal{L}_{\rm eff},\\
         \mathcal{L}_{\rm eff}
         &=\left|\del_t \psi\right|^2
       -  \left|\frac{\partial\psi}{\partial r}\right|^2
       - m^2_\phi (1-2\Psi)
         |\psi|^2-V_{\rm eff}(|\psi|,\omega)
         -\frac{1}{8\pi G}\left(\frac{\partial \Psi}{\partial r}\right)^2,
\end{align}
where $\psi\equiv e^{-i\,\omega\,t}\psi_1$
and we impose a boundary condition $\Psi(r\rightarrow \infty)=0$.
The equations of motion of this system are given by 
\begin{align}
         &\del_t^2\psi-\frac{\del^2\psi}{\del r^2} - \frac{2}{r}\frac{\del\psi}{\del r}
         		+m^2_\phi(1-2\Psi)\psi+\frac{\partial\,V_{\rm eff}}{\partial \,\psi^\dagger} =0 ,\\
         &\frac{1}{4\pi G} \left(\frac{\del^2\Psi}{\del r^2} + \frac{2}{r}\frac{\del\Psi}{\del r}\right)= -2m^2_\phi|\psi|^2.
\end{align}
Note that the time derivative part equivalent to the one obtained from a free field theory,
and the conserved charge $Q$ is defined as usual by Eq. \eqref{Qdef}.

The way to obtain a $Q$-ball (or an axion star) configuration is as follows.
For a given $\omega$, we can define three dimensional action
\begin{align}
         S_3(\omega)\equiv \int_{\bm x}
         \left|{\bm\nabla}\psi\right|^2
    +   \left(m^2_\phi (1-2\Psi)-\omega^2\right)
         |\psi|^2+V_{\rm eff}(|\psi|^2)
	+\frac{\left({\bm\nabla}\Psi\right)^2}{8\pi G}.
\end{align}
Then,
the bounce solution $\psi({\bm x})$ of $S_3(\omega)$ corresponds to the $Q$-ball solution
$e^{-i\omega t}\psi({\bm x})$.
$S_3(\omega)$ can be regarded as a three dimensional action for
two coupled scalar fields $\psi$ and $\Psi$ which have
canonical kinetic terms, and the potential is given by
\begin{align}
         V_{S_3}= \left(m^2_\phi (1-2\Psi)-\omega^2\right)
         |\psi|^2+V_{\rm eff}(|\psi|^2).
\end{align}
The existence of the bounce solution is ensured by the following three conditions
(see~\cite{Blum:2016ipp} for more details):
\begin{itemize}
\item The potential $V_{S_3}$ is non-negative in an open neighborhood which includes the origin $\psi,\Psi=0$;
\item The potential $V_{S_3}$ becomes negative for some $\psi$, $\Psi$;
\item The effective potential $V_{\rm eff}(|\psi|)$ satisfies 
$\lim_{|\psi|\to\infty}\frac{V_{\rm eff}}{|\psi|^6}\geq0$.
\end{itemize}
In general, the solution with the minimal action is ensured to be
$O(3)$ symmetric in space even when there are a lot of fields.
The first condition is satisfied if and only if $m_\phi-\omega>0$.
The second condition is automatically satisfied because
in the large $\Psi$ limit, the potential becomes negative.
The third condition is satisfied as long as the potential is not so pathological.
Note that these conditions do not ensure the validity of the perturbative analysis.
For example, the obtained solution may have 
a large $\Psi = \delta_g\sim\mathcal{O}(1)$.
We need to check the validity of perturbations separately.

The relevant question here is whether the obtained solution is stable
against small perturbations or not.
Even with the gravitational interaction,
stability is ensured provided that
\beq
\frac{\omega}{Q}\frac{\dd Q}{\dd \omega} < 0,
\eeq
which is both a necessary and sufficient condition.
We prove this condition in appendix~\ref{app:stab}. Note that $\omega$ is still given by Eq. \eqref{dEdQ}.

\subsection{Discussion about expansion parameters}

We may roughly estimate the small parameters of the EFT as\footnote{We will ignore $\delta_t$ for the purposes of this section.} 
\begin{align}
 \delta_x &\sim \frac{(\bm{\nabla}\phi)^2}{m_\phi^2\,\phi^2} \sim \frac{1}{(m_\phi\,R)^2}, \\[.5em]
 \delta_V &\sim \frac{\lambda_4\,\phi^4}{m_\phi^2\,\phi^2} \sim \frac{\lambda_4\,\phi^2}{m_\phi^2}, \\[.5em]
 \delta_g &\equiv \Psi \sim \frac{1}{M_P{}^2}\int\,\frac{m_\phi^2\,\phi^2}{r} \dd^3r \sim \frac{m_\phi^2\,\phi^2}{M_P{}^2} R^2,
\end{align}
where $R$ is the size of the oscillon, and we took the self-interaction potential from Eq. \eqref{V4}. The total oscillon mass can be estimated as $M \sim m_\phi^2\,\phi^2\,R^3$, which implies
\begin{equation}
  \delta_x \sim  \frac{1}{(m_\phi\,R)^2}, \qquad
 \delta_V \sim \frac{\lambda_4\,M}{m_\phi^4\,R^3}, \qquad
 \delta_g \sim \frac{M}{M_P{}^2\,R}.
\end{equation}
The size of these parameters determines which EFT corrections become important.

We can build intuition about this by looking at stable bound states. Suppose $\lambda_4<0$ (attractive force). Then a bound state can be supported by a balance of the kinetic pressure against the gravitational and self-interaction terms in the equation of motion (this is the standard picture of a boson star). If $\delta_V \ll \delta_x\sim \delta_g$, then kinetic pressure supports the state against gravity. In this regime we have
\begin{equation}
 M \sim \left(\frac{M_P}{m_\phi}\right)^2\,\frac{1}{R}, \qquad \text{ (gravitating)}
 \label{gravity relation}
\end{equation}
which is the standard mass-radius relation for a boson star with no self-interactions \cite{Kaup:1968zz,Ruffini:1969qy,Breit:1983nr}. On the other hand, if $\delta_g \ll \delta_V \sim \delta_x$, then we have 
\begin{equation}
 M \sim \frac{m_\phi^2}{|\lambda_4|}\,R, \qquad \text{ (self-interacting)}
\end{equation}
which is the mass-radius relation for non-gravitating boson stars \cite{Chavanis:2011zi,Chavanis:2011zm}. These non-interacting (non-gravitating) states are stable (unstable) under perturbations. If all three terms are of similar order $\delta_x \sim \delta_V \sim \delta_g$, we have a transition region which includes a critical (maximum) mass at
\begin{equation}
 M_{c} \sim \frac{M_P}{\sqrt{|\lambda_4|}},
\end{equation}
corresponding to a radius of
\begin{equation}
 R_c \sim \frac{\sqrt{|\lambda_4|}\,M_P}{m_\phi^2}, 
 \label{critical radius}
\end{equation}
reproducing the standard results \cite{Chavanis:2011zi,Eby:2014fya}.

This raises another important point about the EFT. When we compute corrections at some order in $\epsilon_{x,V,t}$, this corresponds to corrections of size $\delta_{x,V,t}$ at this order. If $\delta_g$ is as large as the other small parameters in the problem (as in a stable boson star), then it is not consistent to neglect corrections at higher order in $\delta_g$. 
Another way to say this is that, if gravity contributes in the equation of motion, then post-Newtonian corrections proportional to $\Psi^2$ in the equations will be as important as those proportional to $\nabla^4\phi$, $\lambda_4^2$, etc., which are the ones calculated in all formulations of scalar EFT we have been discussing \cite{Mukaida:2016hwd,Braaten:2016kzc,Namjoo:2017nia,Eby:2017teq}. 
The application of these EFTs beyond leading order, in a system where gravity contributes in an important way in the equations of motion, is not correct. 
Therefore, 
we need to compute corrections coming from the gravity effect $\epsilon_g$ 
up to the same order with $\epsilon_x$ if we consider the case where the clump forms due to the gravitational interaction. We leave a full analysis of higher-order gravitational corrections to future work.

\subsection{Numerical results}
\label{sec:axion}

\begin{figure} 
   \centering
   \includegraphics[width=4.5in]{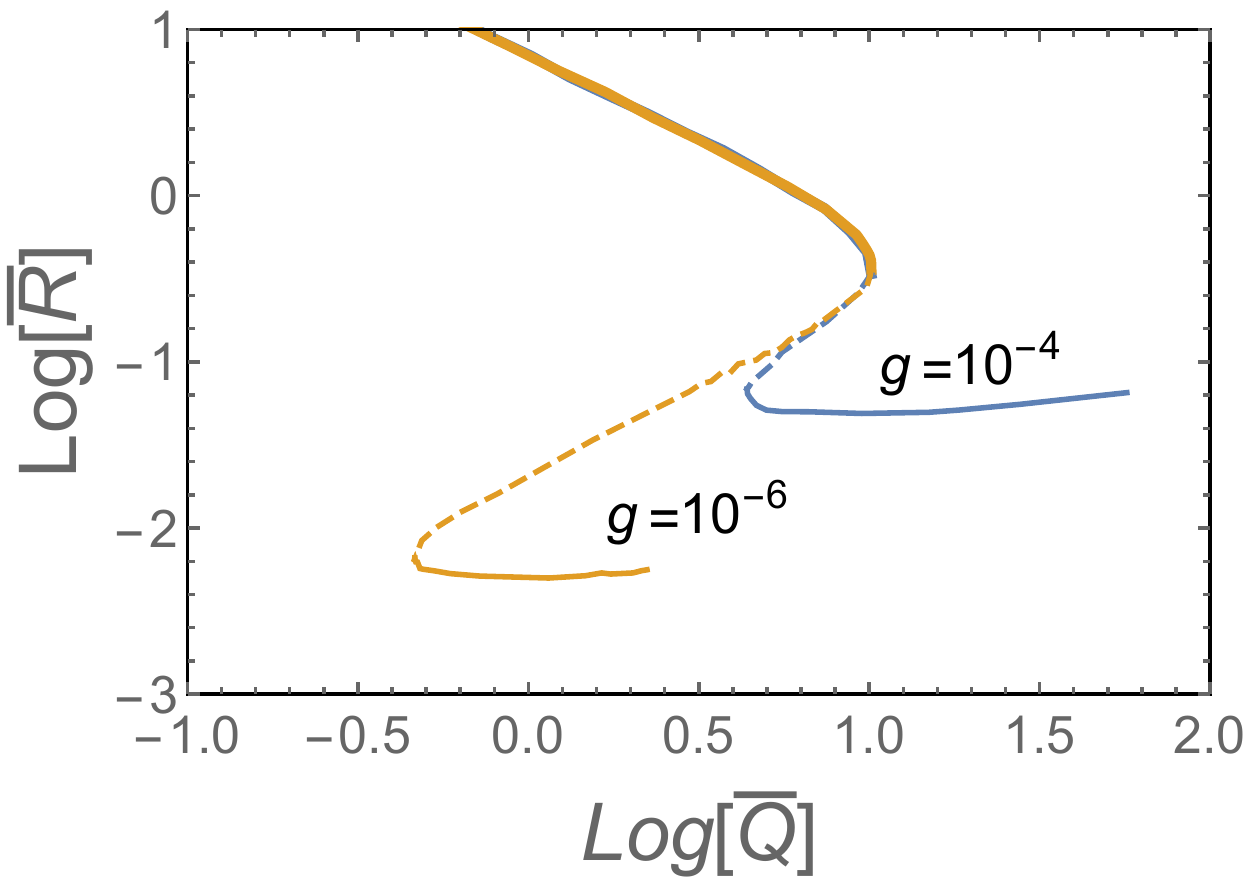} 
   \vspace{0.5cm}\\
    \includegraphics[width=4.5in]{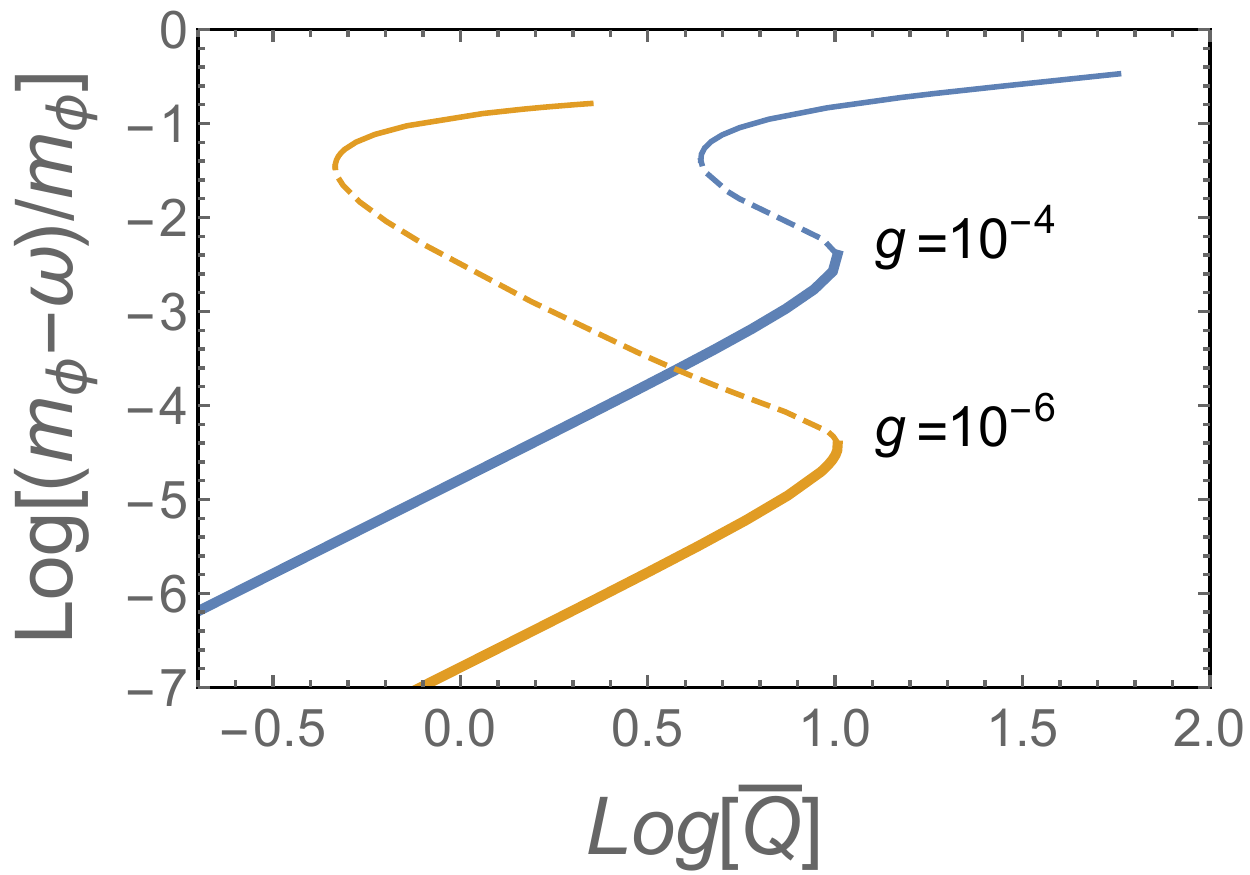} 
   \caption{
   Plots for the radius (upper panel) and $m_\phi - \omega$ (lower panel) as a function of charge for the axion star with $g = 10^{-4}$ (blue line) and $10^{-6}$ (yellow line). 
   The thick lines represent the dilute axion star, 
   the thin lines represent the dense axion star, 
   and the dashed lines represent the unstable transition branch of solutions. 
   }
   \label{fig:phase}
\end{figure}

Here we show a numerical result for an axion star, 
where the interaction potential is given by a relativistic periodic potential minus the mass term,
\begin{equation}
 V_a(\phi) = m_\phi^2\,f_a^2\left[1 - \cos\left(\frac{\phi}{f_a}\right)\right] - \frac{m_\phi^2}{2}\phi^2
 		\approx \frac{\lambda_4}{4!}\phi^4 + \frac{\lambda_6}{6!\,m_\phi^2}\phi^6.
\end{equation}
In general, such a potential will have contributions at every even power of the field $\phi$, but for simplicity we will truncate at $\phi^6$. This is the minimal axion potential
for which a dense axion star solution exists \cite{Eby:2016cnq}.
The coefficients are $\lambda_4 = - \sqrt{\lambda_6} = - (m_\phi/f_a)^2$, 
where $f_a$ is an axion decay constant. In this section, we use the EFT contributions from the $\phi^6$ potential, which we calculate in appendix \ref{app:phi6}, and we take into account $\epsilon_V^2$ as well as $\epsilon_g$ corrections, but neglect higher orders in $\epsilon_x$ and $\epsilon_t$. 

In order to perform numerical simulations, 
it is convenient to rescale variables and fields 
such that they are dimensionless. 
We adopt the following rescaling: 
\beq
 &&\psi(r) = \frac{m_\phi^2}{M_P \abs{\lambda_4}} \bar{\phi}(\bar{r}), 
 \\
 &&r = \frac{\sqrt{\abs{\lambda_4}} M_P}{m_\phi^2} \bar{r}, \label{Rres}
 \\
 &&Q = \frac{M_P}{m_\phi \sqrt{\abs{\lambda_4}}} \bar{Q}, \label{Qres}
 \\
 &&\Psi(r) = g f(\bar{r}) + \frac{1}{2} \lmk 1 - \frac{\omega^2}{m_\phi^2} \rmk, 
\eeq
where we define 
\beq
 g \equiv \frac{m_\phi^2}{M_P^2 \abs{\lambda_4}} = \lmk \frac{f_a}{M_P} \rmk^2. 
\eeq
The equations of motion are given by 
\beq
 &&\bar{\phi}'' + \frac{2}{r} \bar{\phi}' + 2 f \bar{\phi} + \frac12 \bar{\phi}^3 
 - \frac{3g}{32} \bar{\phi}^5 = 0, \label{numEq1}
 \\
 &&f'' + \frac{2}{r} f' + 8 \pi \bar{\phi}^2 = 0, \label{numEq2}
\eeq
where the prime denotes the derivative with respect to $\bar{r}$. 
The boundary conditions are now given by 
\beq
 &&\bar{\phi}' (0) = 0, 
 ~~~~\bar{\phi}(\infty) = 0, 
 \\
 &&f'(0) = 0, 
 ~~~~f(\infty) = \frac{1}{2g} \lmk \frac{\omega^2}{m_\phi^2} - 1 \rmk. 
\eeq

We numerically solve the equations of motion \eqref{numEq1} and \eqref{numEq2}
with various initial conditions. 
Then we calculate the total charge (i.e., the total particle number $Q$) 
and the radius $R$ of the configuration. 
Note that we define the radius as the one at which $90\%$ of the energy is enclosed. 
The resulting phase diagram is shown in Fig.~\ref{fig:phase}, 
where we take $g = 10^{-4}$ and $10^{-6}$. In the plot, we show the rescaled quantities $\bar{R}$ and $\bar{Q}$, defined by the rescaling of Eqs. \eqref{Rres} and \eqref{Qres}

For a large radius, the gravity effect dominates 
and we obtain $R \propto 1/Q$, which is consistent with \eq{gravity relation}. 
This branch is shown as thick lines in Fig.~\ref{fig:phase}. 
As $\bar{R}$ decreases and reaches to ${\cal O}(1)$ (see \eq{critical radius}), 
the oscillon enters into the regime of $d Q / d \omega >0$ 
and becomes unstable. 
This is shown as dashed lines in the figure. 
There is another stable branch in the phase diagram, shown as thin solid lines, 
where the radius is of the order of (but is still larger than) $m_\phi^{-1}$. 
The solution in this regime has been well studied in the context of oscillons, 
where the gravity effect is negligible. In the context of axion clumps, 
it is sometimes called a dense axion star \cite{Braaten:2015eeu} or axiton \cite{Kolb:1993zz}. 
A typical radius of an axion star can be estimated as $\bar{R} \sim \sqrt{g / \delta_x}$ 
from the dimensional analysis. 
For the dense axion star, $\delta_x$ is smaller but not much smaller by many orders of magnitude than unity. Therefore we expect $\bar{R} \sim \sqrt{g}$ for the dense axion star, 
which is consistent with our numerical results. 

If gravity is absent, 
the condition for the existence of axion star solutions is given by \eq{conditions}. 
We find that it cannot be satisfied unless $m_\phi-\omega$ is not so suppressed; near the right endpoint of Fig.~\ref{fig:phase},\footnote{
Strictly speaking, we can find a solution for an arbitrary large $Q$. In the limit of $Q \to \infty$, the frequency $\omega$ approaches to some constant value~\cite{Amin:2010jq}. We denote the endpoint
referring to this constant $\omega$.
}
the expansion parameters are not small and the non-relativistic expansion may not be a good approximation.
It is not directly clear how next-to-leading order corrections would affect the endpoint of axion star solutions.
In addition, the decay rate in this regime is not suppressed, as we will see in the next section. 
Our numerical results show that 
it becomes difficult to find a long-lived dense axion star solution 
if we take smaller $g$.

Our results are broadly consistent with the ones in previous literature, as in most works the authors ignore $\epsilon_x$ and $\epsilon_t$ corrections (as we did in this section)~\cite{Braaten:2015eeu,Chavanis:2017loo,Schiappacasse:2017ham}. A leading-order analysis of the first nontrivial harmonic in a dense axion star was performed in \cite{Visinelli:2017ooc}, whereas all leading-order relativistic corrections were taken into account in \cite{Eby:2017teq}.
Here we have taken into account the leading order relativistic corrections at $\mathcal{O}(\epsilon_V^2)$
and our calculation can be applicable to relatively large $\psi_1$, 
which is the case for the dense axion star branch with a large $\bar{Q}$. 
Of course, any leading-order analysis will break down for the very strongly-bound regime of dense axion stars, where $m_\phi - \omega$ is no longer small.
In our analysis, stability against small perturbations is also confirmed in both dilute and dense branches, 
as seen in the lower panel of Fig.~\ref{fig:phase}, which is in very good agreement with the stability analysis of \cite{Eby:2017teq}.

\section{Classical decay rate}
\label{sec:decay}

The number conservation is approximate and is violated by relativistic particle production, 
which leads to a finite lifetime for these localized clumps. 
Their lifetime can therefore be estimated from the imaginary part of the Lagrangian in the EFT, 
which is calculated by cutting relativistic propagators as done in Ref.~\cite{Mukaida:2016hwd}.
An analysis of quantum decay for axion stars was performed by \cite{Braaten:2016dlp}.

Let us first confirm that the imaginary part $\Gamma$ of the action breaks the charge conservation.
Taking a time derivative of \eq{Q}, we readily get
\begin{align}
	\dot Q &= 2 \omega \int_{\bm{x}} 
	\left( \dot \psi_1^\dag \psi_1 + \psi_1^\dag \dot \psi_1 \right) \\
	& \simeq
	\int_{\bm{x}} \left( \psi^\dag_1 \frac{\delta \Gamma}{\delta \psi^\dag_1} + \text{h.c.} \right),
	\label{eq:number_v}
\end{align}
which clearly shows that $\Gamma$ breaks the approximate number conservation.
Note that there is no time dependence on $\psi_1$ for the stable configuration if we neglect $\Gamma$.

\begin{figure}
	\centering
	\includegraphics[width=.40\textwidth]{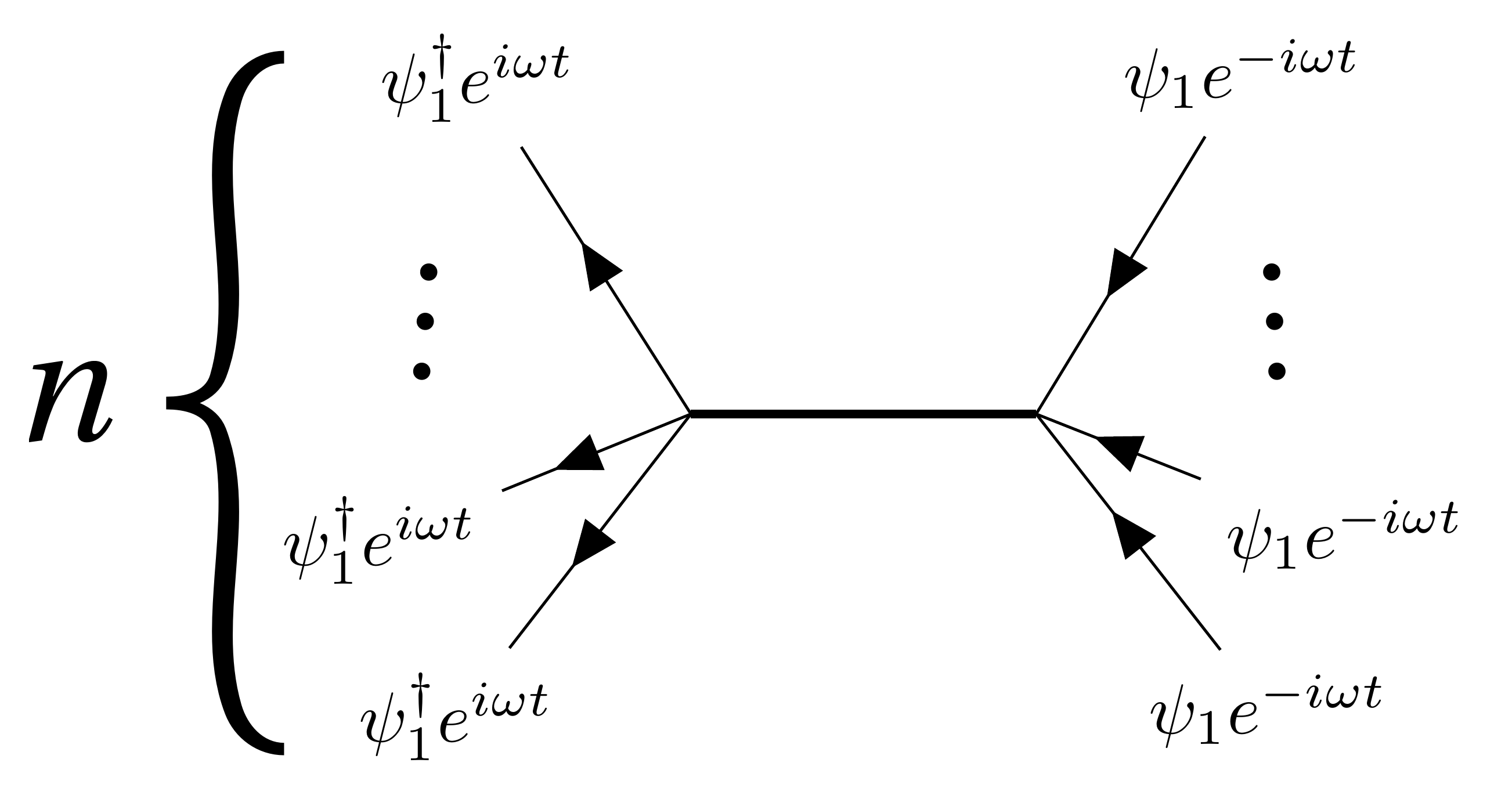}
	\caption{The n to n process mediated by one relativistic particle is shown. The off-shell propagator for the internal line is determined by Eq. \eqref{Skin+mass}.}
	\label{fig:nton}
\end{figure}

Now we are in a position to evaluate how $\Gamma$ depends on $\psi_1$.
Since we focus on a classical NREFT,
the relevant cutting diagrams which yield $\Gamma$ must be tree-level. (We will comment on quantum loop diagrams shortly.)
Motivated by this observation, we consider 
an $n$ to $n$ process mediated by one relativistic particle. See also Fig.~\ref{fig:nton}.
To be concrete, we consider a potential of the form
\begin{equation} \label{Vn}
 V_n(\phi) = \frac{\lambda_{n+1}}{(n+1)! m_\phi^{n-3}}\phi^{n+1}, 
\end{equation}
in the original relativistic theory. After using the EFT expansion of Eq. \eqref{decomposition1}, this term gives 
\begin{align}
	\mathcal{L} \supset \frac{\lambda_{n+1}}{n! m_\phi^{n-3}} \psi_1^n e^{- i n \omega t} \delta \phi + \text{h.c.},
\end{align}
where $\delta \phi$ represents relativistic modes collectively. Classical decays of this type were first considered in the context of axion stars in \cite{Eby:2015hyx}.

The imaginary part appears when the relativistic propagator hits the pole of
\begin{align}
	\int_{x}\frac{\lambda_{n+1}^2}{\left(  n! m_\phi^{n-3} \right)^2} \psi^n_1 
	\frac{1}{ n^2 \omega^2 + \bm{\nabla}^2 - m_\phi^2 + i \epsilon} \psi_1^\dag{}^n,
	\label{eq:cut_nton}
\end{align}
where we have taken $\del_t \psi_1 = 0$ for the stable configuration, and we use the Feynman boundary condition for the relativistic propagator.\footnote{
	Strictly speaking, we would like to study the dynamics of $\psi_1$ and hence it is more appropriate to take the closed-time-path formalism. Then this term looks like
\begin{align}
	\int_{x} \frac{\lambda_{n+1}^2}{2 \left(n ! m_\phi^{n-3} \right)^2}
	\left( \psi{_1^n}^{(+)} - \psi{_1^n}^{(-)} \right) \frac{1}{ (n\omega + i \epsilon )^2 + \bm{\nabla}^2 - m_\phi^2} \frac{1}{2}\left( \psi{{_1^\dag}^n}^{(+)} + \psi{_1^n}^{(-)} \right) + \text{h.c.},
\end{align}
where $\pm$ denotes the fields residing on the upper/lower contour in the closed-time-path formalism as mentioned in Ref.~\cite{Mukaida:2016hwd}. The equation of motion for $\psi_1$ can be derived by taking a derivative with respect to $\psi_1$ and then $\psi_1^{(+)} = \psi_1^{(-)}$. See Ref.~\cite{Mukaida:2016hwd} for more details.
Although the propagator for the relativistic mode here is the retarded one, one can show that the result coincides with Eq.~\eqref{eq:cut_nton} for vacuum of the relativistic mode where $G_\text{ret} = G_\text{Fyn}$ holds. Here $G_\text{ret/Fyn}$ represents the retarded/Feynman propagator.
}
It is clear that, if $\psi_1$ is homogeneous, there is no pole and hence no imaginary part.
Thus, the decay rate strongly depends on how $\psi_1$ localizes in space. This is why we first discuss the profile of $\psi_1$ by neglecting $\Gamma$. This procedure can be justified a posteriori if the decay from $\Gamma$ is much slower than the typical formation time scale of classical lumps.
In other words, if this is the case, one may assume that the scalar field can track the stable solution during the course of its slow decay. Note in addition that this is a fully relativistic phenomenon which cannot be captured by any finite expansion in terms of spatial derivatives of the propagator in Eq. (\ref{eq:cut_nton}).

To see the structure more clearly, we move to the Fourier space. Then 
the imaginary part of the action $\Gamma$ gives rise to a decay rate $\Gamma_{n \to 1} = \int d^3x \Gamma$. Then
Eq.~\eqref{eq:number_v} from this particular diagram can be evaluated as
\begin{align}
	&i \Gamma_{n \to 1} = - \pi i \frac{\lambda_{n+1}^2}{(n! m_\phi^{n-3})^2} 
	\abs{\overline \psi_1}^{2n}
	\int_{\bm{p}} \abs{j_n (\bm{p})}^2 \delta \left( 
	\bm{p}^2 - n^2\omega^2 + m_\phi^2 
	\right), 
\\
	&\dot{Q}_{n \to 1} = 2 n \Gamma_{n \to 1}, 
	\label{eq:nto1}
\end{align}
which was computed in detail in \cite{Mukaida:2016hwd} (though we have chosen a different normalization for the interaction term). We have factorized a typical amplitude of the profile as
$\psi_1 (\bm{x}) = \overline \psi_1 j (\bm{x})$,
and defined
\begin{align}
	j_n (\bm{p}) \equiv \int_{\bm{x}} e^{- i \bm{p} \cdot \bm{x}} \left[j(\bm{x})\right]^n.
	\label{j_n}
\end{align}
Now it is obvious that the classical decay rate is non-zero if $j_n (\bm{p})$ has a non-zero value for $p = \sqrt{n^2 \omega^2 - m_\phi^2}$, which originates from a spatial gradient of the localized lump.

We can calculate $\dot{Q}_{n \to 1}$ numerically by taking a Fourier transformation for the configuration of clump derived from \eq{EoM}. 
One may assume that the profile is approximated by the Gaussian: 
\begin{align}
	\psi_1 (\bm{x}) = \overline{\psi}_1 e^{- r^2/2 R^2},
\end{align}
in which case we can analytically calculate \eq{j_n}: 
\begin{align}
	j_n (\bm{p}) = \left( \frac{2 \pi R^2}{n} \right)^{3/2} 
	e^{- R^2 \bm{p}^2 / 2 n}.
\end{align}
Substituting this into Eq.~\eqref{eq:nto1}, we can evaluate the decay rate: 
\begin{align} \label{Qnto1}
	\dot{Q}_{n \to 1} 
	= -
	\omega\sqrt{n^2-(m_\phi/\omega)^2}
	\frac{32 \pi^5 \lambda_{n+1}^2}{n^2 (n!)^2 } (m_\phi R)^6
	 \lmk \frac{\abs{\overline{\psi_1}}}{m_\phi} \rmk^{2n} 
	 \exp \lkk
	  - \frac{n^2 - m_\phi^2/ \omega^2}{n} \left(R\omega \right)^2 \rkk. 
\end{align}
One can clearly see that the rate is suppressed for $R \omega  \simeq R m_\phi \gg 1$ which is nothing but the non-relativistic condition \eq{small del_x}.
Thus, our treatment is justified a posteriori.
For a larger $n$, it is not easy to produce the relativistic particles because we need more energy to hit the pole.
This is the case of the Gaussian profile, but it is possible to use other profiles to calculate the decay rate. Importantly, the decay rate depends on the tail of the momentum distribution, and so a compact function will give an incorrect result. It is not difficult to use the exact numerical solution of \eq{EoM} for the oscillon profile; this was what was done in \cite{Eby:2015hyx}. The result is parametrically similar to what we have estimated in Eq. \eqref{Qnto1} using the Gaussian ansatz.\footnote{In \cite{Eby:2015hyx}, the expansion parameter was taken to be $\Delta = \sqrt{1 - \omega^2/m_\phi^2} \propto \delta_t$, and the resulting decay rate was proportional to a factor of $\exp\left(-1/\Delta\right)$. With the approximate relation $\delta_x \sim (R m_\phi)^{-2} \sim (R \omega)^{-2}$ and $\delta_t \sim \delta_x$ from the equation of motion, one finds the same exponential dependence $\exp\left(-1/\delta_t\right)$ in Eq. \eqref{Qnto1}. Because this term dominates the behavior of the decay rate, we take this level of agreement to be sufficient.}

One non-trivial example is the case when the profile is rather flat~\cite{Amin:2010jq}.
In such a case, $j_n(p)$ can become zero at a certain radius. As a result,
the configuration stays almost all time at the point where the lowest $j_n$, becomes zero since the decay rate is highly suppressed at that point~\cite{Mukaida:2016hwd}.
 Note in addition that the decay rate is proportional to $\exp[- {\cal O}(\delta_x^{-1})] $
and such effects can not be realized by any order of $\epsilon$ (or $\delta$) expansions.
In this sense, this decay process can be regarded as a kind of non perturbative one.

Here, let us estimate a typical time scale due to this classical decay 
by using a Gaussian profile.
In that case we have $\delta_x\equiv 1/(m_\phi R)^2\ll1$, where $R$ is the radius parameter in the Gaussian profile.
The total number of particles is given by $2Q\simeq 4\pi \sqrt{\pi}R^3 m_\phi|\overline{\psi}_1|^2$.
In addition, we define 
\begin{equation}
\delta_{V_n}\equiv  \frac{\lambda_{n+1} |\overline{\psi_1}|^{n+1}}{(n+1)!m^{n-3}_\phi} \cdot \frac{1}{m_\phi^2|\overline{\psi_1}|^2} \ll 1, 
\end{equation}
where we used the potential $V_n$ of Eq. \eqref{Vn}. Note that $\delta_x\gtrsim \delta_{V_n}$ is expected from the equation of motion.
Then, we have the following expression for the typical time scale for decay:
\begin{align} \label{timescale}
       1/T_{\rm typ}\equiv  \left|\frac{\dot{Q}}{Q}\right|_{n\to1}
       		\simeq
         \frac{4\pi^4}{n^2}\sqrt{\frac{n^2-1}{\pi}}\sqrt{\delta_{V_n}}\left(\frac{\delta_{V_n}}{\delta_x}\right)^{3/2}\exp\left(-\frac{n^2-1}{n\delta_x}\right) \times m_\phi.
\end{align}
If $\delta_x$ is not so suppressed, this decay process may determine the fate of
the configuration.

Now we shall consider the axion star, which is numerically calculated in Sec.~\ref{sec:axion}. 
In this case, 
the typical value of $\delta_x$ can be estimated from the top panel Fig.~\ref{fig:phase} 
because 
\beq
 \delta_x 
 &\sim& \frac{1}{(m_\phi R)^2} 
 \\
 &=& \frac{\lambda_4 M_P^2}{ (m_\phi \bar{R})^2 }
 = \frac{g}{\bar{R}^2}. 
\eeq
Using $\delta_x \sim \delta_t = \abs{m_\phi - \omega}/ m_\phi$, 
one can also read it from the bottom panel of the figure. 
In the dense axion star branch, it is easy to see that $\delta_t$ is of order $0.1-1$ 
and hence by examining Eq. \eqref{timescale}, we expect that these objects would decay before the present epoch. This was concluded also in the analysis of \cite{Eby:2015hyx,Visinelli:2017ooc}.
On the other hand, in the dilute axion branch, 
$\delta_t$ is many orders of magnitude smaller. We can see in Figure \ref{fig:phase} that the ratio of $\delta_t$ in the dilute branch compared to that of the dense branch is smaller than $g = (f_a / M_P)^2$. 
Since we expect $g \sim 10^{-6}$ for the GUT-scale axion 
or $g \sim 10^{-14}$ for the (standard) QCD axion, 
the lifetime of dilute axion star is much longer than the present age of the universe 
because of the exponential factor in the decay rate. 
Therefore we conclude that dilute axion stars can survive until the present, 
if they formed in the early universe. The decay rate can be significantly larger when $g$ is larger, as in models of Fuzzy Dark Matter \cite{Eby:2017azn}.

Finally, we comment on quantum decay of oscillons. 
In this work, we have discussed the classical decay of oscillons via relativistic corrections, where 
the gradient term hits the pole of the propagator in the tree-level diagram. 
One may calculate loop diagrams to take into account quantum corrections in the EFT. 
Then there are some corrections to the imaginary term of the EFT 
from the cutting diagrams, which represent quantum decay of oscillon. 
If the decay rate is sufficiently small, 
it can be roughly estimated as the elementary decay rate of scalar field times the number of particles inside the oscillon. 
However, if the decay rate is sufficiently large, 
the statistics of daughter particles may become relevant. 
If the daughter particle is fermionic and obeys Fermi-Dirac statistics, 
its production rate from the surface of the oscillon has an upper bound by the Pauli exclusion principle~\cite{Cohen:1986ct, Kawasaki:2012gk}. 
If the decay rate is saturated, it is proportional to the number of degrees of freedom for particles that interact with the oscillon. 

On the other hand, if the daughter particle is bosonic and obeys Bose-Einstein statistics, 
its production rate may be enhanced by the Bose-enhancement effect~\cite{Hertzberg:2010yz, Kawasaki:2013awa}. 
The latter effect may lead to an interesting observable effect for axion star~\cite{Hertzberg:2018zte}. 
The classical decay discussed in this paper could in principle have been affected by the Bose-enhancement effect. 
However, the effect is relevant only
if the production rate is sufficiently larger than the escape rate of daughter particles. 
Therefore this enhancement effect is not important for the classical decay via the self-interaction.

\section{Comparison with other works}
\label{sec:comparison}

In this section we will discuss other methods for calculating relativistic corrections in NREFT.

\subsection{Method proposed by Namjoo, Guth, and Kaiser}

In Ref.~\cite{Namjoo:2017nia}, NGK
proposed an EFT by defining the scalar field with a nonlocal operator, giving rise to an equation of motion with only first-order time derivatives. 
In this case, the denominator of the propagator is linear in the energy, 
resulting in propagation which is only forward in time. 
In other words, in their EFT one integrates out part of the non-relativistic mode with frequency $\sim -m_\phi$.
On the other hand, our EFT contains the second-order time derivatives and integrates out only modes with high frequencies $> \abs{m_\phi}$. Though the results agree at the level of T- and S-matrix elements \cite{Namjoo:2017nia,Braaten:2018lmj}, our method has the advantage of increased computational efficiency and smaller corrections beyond the leading order.
In this section, we consider a polynomial potential with $\phi^4$ and $\phi^6$ order terms, although in \cite{Namjoo:2017nia} only $\phi^4$ was considered. For the purposes of comparison, the relevant $\phi^6$ corrections in the MTY method of the EFT are shown in appendix~\ref{app:phi6}. 

\subsubsection{Canonical transformation}

The Hamiltonian of the relativistic theory (\ref{V2}) is given by 
\beq
 {\cal H} = \frac{1}{2} \pi^2 + \frac{1}{2} \lmk \bm{\nabla} \phi \rmk^2 
 + \frac12 m_\phi^2 \phi^2 + \frac{\lambda_4}{4 !} \phi^4 + \frac{\lambda_6}{6! m_\phi^2} \phi^6, 
\eeq
where $\pi$ is the canonical momentum. 
The canonical variables are $\phi$ and $\pi$. The Hamiltonian can be rewritten in terms of new canonical variables $\psi$ and $i \psi^*$ 
by the following canonical transformation: 
\beq
 \phi (t, {\bm x}) = \frac{1}{\sqrt{2m_\phi}} {\cal P}^{-1/2} 
 \lkk e^{-i m_\phi t} \psi (t, {\bm x}) + e^{im_\phi t} \psi^* (t, {\bm x}) 
 \rkk, 
 \label{trans2}
 \\
 \pi (t, {\bm x}) = -i \sqrt{ \frac{m_\phi }{2}} {\cal P}^{1/2} 
 \lkk e^{-im_\phi t} \psi (t, {\bm x}) - e^{im_\phi t} \psi^* (t, {\bm x}) 
 \rkk. 
 \label{trans}
\eeq
where ${\cal P}$ is defined by 
\beq
 {\cal P} \equiv \sqrt{1- \frac{\bm{\nabla}^2}{m_\phi^2}}. 
\eeq
This should be included since otherwise the equation of motion of $\psi$ 
depends on the spatial derivative of $\psi^*$~\cite{Namjoo:2017nia}. Note for future reference that the dimension of $\psi$ in the normalization of NGK differs from that of Eq. \eqref{decomposition1} by a factor of $\sqrt{m_\phi}$.

This canonical transformation can be done by the following generating function $F[\phi, \psi, t]$: 
\beq
 F[\phi , \psi, t] = i m_\phi  \phi^2 - i \sqrt{2m_\phi } e^{-im_\phi t} \psi \phi + \frac{i}{2} e^{-2im_\phi t}\psi^2, 
\eeq
which satisfies 
\beq
 &&\frac{\del F[\phi , \psi, t]}{\del \phi} = \pi, 
 \\
 &&- \frac{\del F[\phi , \psi, t]}{\del \psi} = i \psi^*. 
\eeq
Hence the new Hamiltonian ${\cal H}'$ is given by 
\beq 
 {\cal H}' [\psi, \psi^*] 
 &=& {\cal H}[\phi, \pi] + \frac{\del F[\phi , \psi, t]}{\del t}
 \\
 &=& m_\phi  \psi^* \lmk {\cal P} - 1 \rmk \psi 
 + \frac{\lambda_4}{4!} \phi^4[\psi, \psi^*] 
 + \frac{\lambda_6}{6! m_\phi^2} \phi^6[\psi, \psi^*], 
\eeq
where $\phi$ and $\pi$ should be rewritten in terms of $\psi$ and $\psi^*$ by using Eqs. \eqref{trans2} and \eqref{trans}. 
This is similar to the Hamiltonian in the non-relativistic field theory if we expand ${\cal P}$ by using $\bm{\nabla}^2 /m_\phi^2 \ll 1$.

The Lagrangian is therefore given by 
\begin{align}
\label{lag1}
         \mathcal{L}=\frac{i}{2}\left(
         \dot{\psi}\psi^*-\psi\dot{\psi}^*
         \right)
         -m_\phi  \psi^* ( {\cal P} - 1) \psi 
         - \frac{\lambda_4}{4!} \phi^4[\psi, \psi^*] 
         - \frac{\lambda_6}{6! m_\phi^2} \phi^6[\psi, \psi^*]. 
\end{align}
As a result, the equation of motion looks like a Schr\"odinger equation 
in quantum mechanics. 
Since we have just used the canonical transformation, 
the commutation relations for $\psi$ and $i \psi^*$ are 
the same as those for $\phi$ and $\pi$. 
Up to now, the calculations are exact.

\subsubsection{Expansion scheme}

Following Ref.~\cite{Namjoo:2017nia}, 
we decompose $\psi(t,{\bm x})$ as
\begin{align}
         \psi (t,{\bm x})=\sum_{\nu=0,\pm1,\pm2,...}\psi_\nu(t,{\bm x})e^{i\nu m_\phi t},
\end{align}
where each of the $\psi_\nu({\bm x},t)$ is assumed to be slowly varying.
The modes with odd $\nu$ do not play any role in the following analysis 
because of the $\mathbb{Z}_2$ symmetry in the original Lagrangian. 

Here, we note that 
the zero point energy is shifted by an amount of $m_\phi $ 
by the canonical transformation. 
Hence the $\psi_0$ represents the non-relativistic field whose kinetic energy is much smaller 
than its rest mass. 
Its rest energy $m$ is removed from the Hamiltonian by the constant shift. 
So one may expect that 
the non-relativistic Lagrangian for $\psi_0$ can be constructed by integrating out $\psi_\nu$ with $\nu \ne 0$. 

However, we should note that the modes $\psi_\nu$ with $\nu \ge 2$ represents 
components of the original scalar field that have negative frequencies. 
In particular, the mode $\psi_2$ satisfies $i  \dot{\psi}_2 + m_\phi \psi_2 = -  m_\phi \psi_2$, where a factor of $m_\phi$ comes from the constant shift of the Hamiltonian.
This means that the EFT of $\psi_0$ includes a contribution from integration of the mode $\psi_2$ with frequency of order $-m_\phi$. This leads to a large correction at the first nontrivial order.

To integrate out relativistic modes, 
we introduce a pseudo  parameter $\epsilon$ and expand $\psi_\nu$ ($\nu \ne 0$) as 
\beq
 \psi_\nu (t, {\bm x}) = \sum_{n=1}^\infty \epsilon^n \psi_\nu^{(n)} (t, {\bm x}). 
\eeq
We assume that $\psi_0$ is the zeroth order for $\epsilon$. 
Then the equation of motion can be expanded by this small parameter. 
The result is given by 
\beq
 &&i \dot{\psi}_0 
 = ({\rm leading \  terms}) + 
 \frac{3\lambda_4}{4! m_\phi^2} 
 \lkk \psi_0^2 \psi_2 
 + 2 \psi_0^* \psi_0 \psi_2^* 
 + \psi_0^{*2} \psi_{-2} 
 + \psi_0^{*2} \psi_4^* 
 \rkk
 \nn
 &&~~~~~~~~~~~~~~~~+
 \frac{15}{4} \frac{\lambda_6}{6! m_\phi^5} 
 \lkk \psi_0^4 \psi_4 
 + \psi_0^4 \psi_{-2}^* 
 + 4 \psi_0^{*} \psi_0^3 \psi_{2} 
 + 6 \psi_0^{*3} \psi_0 \psi_{2}^* 
 \right.
 \nn
 &&~~~~~~~~~~~~~~~~~~~~~~~~~~~~~~~~~~~~~~ \left. + 4 \psi_0^{*3} \psi_0 \psi_{-2} 
 + 4 \psi_0^{*3} \psi_0 \psi_{4}^* 
 + \psi_0^{*4} \psi_{-4} 
 + \psi_0^{*4} \psi_{6}^* 
 \rkk, 
 \label{eff2NGK}
\eeq
at the first order of $\epsilon$ ($=1$). 
Here and hereafter, we set ${\cal P} = 1$ for simplicity and focus only on the effective potential. 
The relativistic modes can be rewritten in terms of $\psi_0$ 
by using $\dot{\psi}_\nu \ll m_\phi \abs{\psi_\nu}$: 
\beq
 &&\psi_{-4} = 
  \frac{1}{4} c_6 \psi_0^5,
 \qquad \qquad
 \psi_{-2} = 
 \frac{1}{2} c_4 \psi_0^3
 + \frac{5}{2} c_6 \psi_0^{*} \psi_0^4,
 \quad
 \\
 &&\psi_{2} = 
 -\frac{3}{2} c_4 \psi_0^{*2} \psi_0
 -5 c_6 \psi_0^{*3} \psi_0^2,
 \\
 &&\psi_{4} = 
 - \frac{1}{4} c_4 \psi_0^{*3}
 - \frac{5}{4} c_6 \psi_0 \psi_0^{*4},
 \qquad \qquad
 \psi_{6} = 
 - \frac{1}{6} c_6 \psi_0^{*5}, 
\eeq
where we define 
\beq
 c_4 = \frac{\lambda_4}{4! m_\phi^3}, 
 \qquad
 c_6 = \frac{3}{4} \frac{\lambda_6}{6! m_\phi^6}. 
\eeq
Substituting these into \eq{eff2NGK}, 
we obtain the effective potential as 
\beq
 V_{\rm eff}(\abs{\psi_1})\big\vert_{\epsilon_V^2}
 &=& 
  \frac{\lambda_4}{(2!)^2} \lmk \frac{\abs{\psi_0}^2}{2m_\phi} \rmk^2 
 + \frac{\lambda_6 - 17/8 \lambda_4^2}{(3!)^2m_\phi^2} \lmk \frac{\abs{\psi_0}^2}{2m_\phi} \rmk^3 
 \nn
 &&~~~~ -11 \frac{\lambda_4 \lambda_6}{(4!)^2m_\phi^4} \lmk \frac{\abs{\psi_0}^2}{2m_\phi} \rmk^4 
 - \frac{131}{6}\frac{\lambda_6^2}{(5!)^2m_\phi^6} \lmk \frac{\abs{\psi_0}^2}{2m_\phi} \rmk^5.  
 \label{eff3}
\eeq
This is different from \eq{eff2} even if we take into account the difference of the normalization of $\psi$ ($\abs{\psi_0}^2/2m_\phi \leftrightarrow \abs{\psi_1}^2$). 
In particular, the signs of first order corrections in NGK and MTY are the opposites of each other. 

However, if we do not integrate out $\psi_2$, 
we obtain 
\beq
 i \dot{\psi}_0 = ({\rm interaction \ terms \ with\ } \psi_2) + \tilde{V}_{\rm eff} (\psi_0). 
\eeq
We found that 
the effective potential $\tilde{V}_{\rm eff}$ is equivalent to \eq{eff2} when we take into account the difference of the normalization of $\psi$. 
We expect that the form of the equation of motion can be factorized into a combination of $\psi_0$ and $\psi_2$, and then the resulting effective potential is given by the same form as $\tilde{V}_{\rm eff}$ and given by \eq{eff2}. 
This is consistent with the fact that 
$\psi_2$, which is a fast oscillating mode from the standpoint of NGK, is a non-relativistic mode from the standpoint of MTY.\footnote{As this paper was being finalized, an update to \cite{Namjoo:2017nia} appeared which contained a discussion of some of these issues, and where the authors performed a comparison of these methods as well.}

\subsection{Method proposed by Braaten, Mohapatra, and Zhang}

Here we explain the method proposed by BMZ in Ref.~\cite{Braaten:2016kzc,Braaten:2018lmj}. 

They first write down the effective Lagrangian 
\beq
 {\cal L}_{\rm eff} = 
\frac{i}{2}\left(
         \dot{\psi}\psi^*-\psi\dot{\psi}^*
         \right)
         - \frac{1}{2m_\phi} \bm{\nabla} \psi^* \cdot \bm{\nabla} \psi 
         - \sum_{n=2}^\infty \frac{\lambda_{2n}'}{(n !)^2} \lmk \frac{\psi^* \psi}{2 m_\phi} \rmk^n. 
\eeq
The effective coupling constants $\lambda_{2n}'$ are determined by 
the following procedure. 
First, they calculate the scattering amplitudes which have only non-relativistic particles in the external lines 
by using the original relativistic Lagrangian 
and the effective Lagrangian. 
Then compare the results and determine $\lambda_{2n}'$ so that the results are consistent. 
Note that the vertex factor in the EFT is given by $\lambda_{2n}' / 2^n$. 
In addition, there should be an additional factor $\sqrt{2}^m$ for the diagram with $m$ external legs in the EFT 
to compare the amplitude with the one calculated by the original theory.

It is important that the contribution to the amplitude 
from the non-relativistic internal line in the EFT 
has a different meaning from the one calculated in the original theory. 
This is because in the non-relativistic EFT 
the denominator of the propagator is linear in the energy, 
resulting in propagation which is only forward in time. 
The resulting EFT 
is therefore equivalent to the one calculated in Ref.~\cite{Namjoo:2017nia}. 
In fact, 
we have checked that the effective potential calculated in Ref.~\cite{Braaten:2016kzc} (i.e. \eq{eff3}) 
is consistent with the one calculated in Ref.~\cite{Namjoo:2017nia} 
at least for the coefficients of $\lambda_4^2$, $\lambda_4 \lambda_6$, 
and $\lambda_6^2$.

\subsection{Consistency check}
\label{sec:consistency}

Here we check the consistency of EFTs: \cite{Mukaida:2016hwd}(MTY) and \cite{Braaten:2016kzc, Namjoo:2017nia,Braaten:2018lmj}(BMZ-NGK).  
Since each EFT defines the dominant non-relativistic part in a different way,
we should be careful about how to compare different EFTs.
One way may be to compare some physical quantities predicted by EFTs which are truncated 
at the same $\epsilon^n$ order.
For example, EFT can predict the frequency-dependent total energy of oscillons $E(\omega)$, which 
is an important physical quantity.
We expect the difference of the physical quantities to be $\mathcal{O}(\epsilon^{n+1})$.

Below, we will show such a consistency check of the effective theory at $\mathcal{O}(\epsilon_V^3)$ 
in \cite{Mukaida:2016hwd} (MTY) and \cite{Braaten:2016kzc, Namjoo:2017nia,Braaten:2018lmj} (BMZ-NGK). 
We consider a simple setup where there is only the quartic coupling and 
no spatial dependence 
for simplicity. 
The Lagrangian and equation of motion are then given by 
\begin{align}
         & -\mathcal{L}=\frac{1}{2}\phi(\partial_t^2+m_\phi^2)\phi+\frac{\lambda_4}{4!}\phi^4,\\
         & \ddot{\phi}+m_\phi^2\phi + \frac{\lambda_4}{3!}\phi^3=0,
\end{align}
where the dot denotes derivative with respect to time.
If the amplitude of $\phi$ is given  by $a\times m_\phi$
with $a$ being some dimensionless number,
the frequency of oscillation $\omega$ can be written as
\begin{align}
\label{omom}
         (\omega/m_\phi)&=\frac{\pi}{2}\frac{\sqrt{1+(\lambda_4/3!) a^2}}{K(k^2)},\\
         k^2&=\frac{(\lambda_4/3!) a^2}{2(1+(\lambda_4/3!) a^2)},
\end{align}
where $K(k^2)$ is elliptic integral of the first kind and given by
\begin{align}
         K(k^2)=\frac{\pi}{2}\sum_{n=0}^{\infty} \left(
         \frac{(2n-1)!!}{(2n)!!}
         \right)^2 k^{2n}.
\end{align}
Note that in the limit $\lambda_4\rightarrow 0$, we recover $\omega/m_\phi=1$.

For this system, we can use each EFT to estimate $\omega$.
The procedure is as follows.
First, we fix the amplitude of the dominant mode ($\psi_1$ in MTY
and $\psi_0$ in BMZ-NGK) to some constant value.
Then, by using the equations of motion derived from each EFT,
we can estimate $\omega$.
On the other hand, we can also estimate the amplitude $a$
which is modified by fast oscillating modes.
Then, we can use the frequency formula above~Eq.~(\ref{omom})
to estimate $\omega$.
If $\omega$ derived from an EFT and the one from the frequency formula
are the same, the EFT can be regarded as a consistent one.

Below, we estimate $\omega$ up to third order in the $\lambda_4$ expansion
by using formula Eq.~(\ref{omom}) and EFT in MTY and BMZ-NGK.
We will see that both EFTs can reproduce the correct result at least to third order in $\lambda_4$.
This means that if we make a prediction of amplitude dependent frequency $\omega(a)$
by using both EFTs at $\epsilon^3$,
the difference appears at $\epsilon^4$.
This fact supports the consistency of both EFTs.

\subsubsection{MTY}
We fix the amplitude of the non-relativistic field to be $|\psi_1|/m_\phi=1$
in order to set $a= 1+\mathcal{O}(\lambda_4)$.
This choice is just
for representational simplicity.
For any value of $|\psi_1|$, we can do the same procedure
and see consistency.

At the order of $\mathcal{O}(\lambda_4^2)$, there exist the following fast oscillating modes:
\begin{align}
         \frac{\phi - [\psi_1e^{-i\omega t}+\text{h.c.}]}{m_\phi} &
         =\left(\frac{(\lambda_4/3!)}{32}\abs{\frac{\psi_1}{m_\phi}}^3
         -\frac{9(\lambda_4/3!)}{128}((\omega/m_\phi)-1)\abs{\frac{\psi_1}{m_\phi}}^3+\frac{3(\lambda_4/3!)^2}{512}\abs{\frac{\psi_1}{m_\phi}}^5
         \right)\nonumber \\
         &\times
         \cos (3\omega t)\nonumber \\
         &+\frac{(\lambda_4/3!)^2}{1024}\abs{\frac{\psi_1}{m_\phi}}^5
         \cos (5\omega t)
         +\mathcal{O}({\lambda^3_4}).
\end{align}
As a result, when we evaluate the expression at $|\psi_1/m_\phi|=1$, the total amplitude is shifted as
\begin{align}
         a&=1+\frac{(\lambda_4/3!)}{32}-\frac{9(\lambda_4/3!)}{128}((\omega/m_\phi)-1)
         +\frac{7(\lambda_4/3!)^2}{512}+\mathcal{O}(\lambda^3_4).
\end{align}
Then, the frequency formula Eq.~(\ref{omom}) gives
\begin{align}
\label{om:y}
         \omega/m_\phi=1+\frac{3}{8}(\lambda_4/3!)-\frac{15}{256}(\lambda_4/3!)^2
         +\frac{123}{8192}(\lambda_4/3!)^3
         +\mathcal{O}{(\lambda^4_4)}.
\end{align}
On the other hand, our EFT gives
\begin{align}
         2(\omega/m_\phi-1)+(\omega/m_\phi-1)^2&=
         \frac{3(\lambda_4/3!)}{4}\abs{\frac{\psi_1}{m_\phi}}^2
         +\frac{3(\lambda_4/3!)^2}{128}\abs{\frac{\psi_1}{m_\phi}}^4
          \nonumber \\
        &-\frac{27(\lambda_4/3!)^2}{512}((\omega/m_\phi)-1)\abs{\frac{\psi_1}{m_\phi}}^4
         +\frac{3(\lambda_4/3!)^3}{512}\abs{\frac{\psi_1}{m_\phi}}^6
         +\mathcal{O}(\lambda^4_4).
         \label{MYT_cons}
\end{align}
Note here that the term proportional to $(\omega/m_\phi - 1)$ comes from the interaction at $\epsilon_V^2 \epsilon_t$.
This is exactly the same term pointed out in Ref.~\cite{Braaten:2018lmj}.
We will also clarify this point in Sec.~\ref{sec:timederivative}.
One may readily solve this equation at $|\psi_1/m_\phi| = 1$ to give
\begin{align} \label{omegaMTY}
         \omega/m_\phi=1+\frac{3}{8}(\lambda_4/3!)
         -\frac{15}{256}(\lambda_4/3!)^2+\frac{123}{8192}
         (\lambda_4/3!)^3
         +\mathcal{O}{(\lambda^4_4)},
\end{align}
which is consistent with the exact result given in Eq. \eqref{om:y}.
\subsubsection{BMZ-NGK}

We fix $|\psi_0|/m_\phi^{3/2}=1/\sqrt{2}$ in order to set $a=1+\mathcal{O}(\lambda)$.
As before, this choice is just for representational simplicity.
At $\mathcal{O}(\lambda^2_4)$, there are $\nu=-4,-2,2,4,6$ modes in $\psi$
(Note that some of these modes are mixed with the non-relativistic mode in MTY sense).
Taking their contributions into account, the total amplitude becomes
\begin{align}
         a=1-\frac{5(\lambda_4/3!)}{32}+\frac{3(\lambda_4/3!)}{128}((\omega/m_\phi)-1)
         +\frac{91(\lambda_4/3!)^2}{1024}
         +\mathcal{O}(\lambda^3_4).
\end{align}
Then, the frequency formula Eq.~(\ref{omom}) gives
\begin{align}
\label{om:g}
         \omega/m_\phi=1+\frac{3}{8}(\lambda_4/3!)-\frac{51}{256}(\lambda_4/3!)^2
         +\frac{1419}{8192}(\lambda_4/3!)^3
         +\mathcal{O}{(\lambda^4_4)}.
\end{align}
On the other hand, this EFT predicts
\begin{align}
         (\omega/m_\phi-1)
         =\frac{3(\lambda_4/3!)}{8}-\frac{51(\lambda_4/3!)^2}{256}
         +\frac{81(\lambda_4/3!)^2}{1024}(\omega/m_\phi-1)
         +\frac{147(\lambda_4/3!)^3}{1024}
         +\mathcal{O}(\lambda^4_4).
\end{align}
Now one can solve this equation easily
\begin{align} \label{omegaNGK}
         \omega/m_\phi=1+\frac{3}{8}(\lambda_4/3!)-\frac{51}{256}(\lambda_4/3!)^2
         +\frac{1419}{8192}(\lambda_4/3!)^3
         +\mathcal{O}{(\lambda^4_4)},
\end{align}
which is consistent with the exact result given in Eq.~(\ref{om:g}).

Finally let us comment on the size of corrections.
Since
MTY and BMZ-NGK are based on different expansion schemes,
the sizes of fast oscillation modes are also different.
The size of fast oscillation modes are estimated as a ratio of
the strength of the source and the size of the propagator.
In general,
the size of the propagator in MTY is larger than that in BMZ-NGK.
This is in part because the size of the propagator in MTY is proportional to $n^2-1$
with $n\omega$ being the frequency of fast oscillation modes while
that of BMZ-NGK is given by $n$.
As a result, corrections to the effective action in MTY becomes
smaller than that in BMZ-NGK.
For example, comparing the size of corrections in Eq. \eqref{omegaMTY} and Eq. \eqref{omegaNGK}, we can see that at higher orders, the corrections become increasingly large in the method of NGK compared to MTY.
This is one advantage supporting the use of the method of MTY, whose higher order corrections
are relatively small.

\subsubsection{On interaction with time derivative}
\label{sec:timederivative}

In Ref.~\cite{Braaten:2018lmj}, they argued that there is an error in MTY \cite{Mukaida:2016hwd}, namely there is a missing term with a time derivative.
However, as we have seen in Sec.~\ref{sec:consistency}, both EFTs give the same result, which suggests their claim is incorrect.
In this section, we clarify that our EFT contains the required term.
We explicitly show that the interaction with a time derivative arises if one would like to match the EFT with scatterings of NR free particles as done in Ref.~\cite{Braaten:2018lmj}.

First, we illustrate how to derive the interaction term proportional to $(\omega - m_\phi)$ in the case of the stationary $\psi_1$ in Sec.~\ref{stable}.
Let us start from Eq.~\eqref{3to3}. Thanks to $\partial_t \psi_1 = 0$, one finds
\begin{align}
	- \int_x \frac{\lambda_4^2}{(3 !)^2} \psi_1^3
	\frac{1}{9 \omega^2 - m_\phi^2 + \bm{\nabla}^2 + i \epsilon} \psi_1^\dag{}^3
	= - \int_x \frac{\lambda_4^2}{8 (3 !)^2} 
	\left( 1 - \frac{9 (\omega / m_\phi - 1 )}{4} \right)\frac{\abs{\psi_1}^6}{m_\phi^2} + \cdots.
	\label{time_dep_1}
\end{align}
The second term represents the correction at $\epsilon_V^2 \epsilon_t$ which gives the term proportional to $(\omega - m_\phi)$ in Eq.~\eqref{MYT_cons}.

Now let us confirm that the same term can be reproduced even if we take $\omega \to m_\phi$ but keep $\partial_t \psi_1 \neq 0$ as done in Ref.~\cite{Mukaida:2016hwd}. This limit is useful if one would like to compute the scattering amplitude of free NR particles rather than to obtain the stationary solution.
One easily gets
\begin{align}
	- \int_x \frac{\lambda_4^2}{(3 !)^2} \psi_1^\dag{}^3 \frac{1}{8 m_\phi^2 + 6 i m_\phi \partial_t + \bm{\nabla}^2 - \partial_t^2 + i \epsilon} \psi_1^3
	= - \int_x \frac{\lambda_4^2 \abs{\psi_1}^4}{8 (3 !)^2 m_\phi^2} 
	\left(\abs{\psi_1}^2 - \frac{9}{8}\frac{\psi_1^\dag i \overleftrightarrow{\partial_t} \psi_1}{m_\phi} \right) +  \cdots.
	\label{time}
\end{align}
The second term is nothing but the one pointed out in Ref.~\cite{Braaten:2018lmj},
which is clearly included in our EFT.
We can also see that this expression is equivalent to Eq.~\eqref{time_dep_1} if one replaces $\psi_1 (t, \bm{x})$ with $e^{- i (\omega - m_\phi)t} \psi_1 (\bm{x})$.

\section{Conclusion} \label{sec:Conclusions}

We have provided a method to construct a classical NREFT in a scalar field theory including the effect of gravity. 
There are several advantages in our EFT:

\begin{itemize}
\item the calculation is easy and straightforward by using Feynman diagrams; 
\item the reason of (approximate) stability is clear; 
\item it is easy to calculate the background configuration; 
\item it is straightforward to include the effect of gravity; 
\item the lifetime can be calculated from the imaginary part of the Lagrangian in the NREFT;
\item the higher order corrections become relatively small. 
\end{itemize}
We have clarified an expansion scheme 
for gravitational corrections as well as relativistic corrections. 
Since the gravitational potential has a radial dependence, 
we need to solve coupled differential equations to determine the oscillon profile. 

Since the number of particles is approximately conserved in the NREFT, 
the oscillon can be understood as a projection of a Q-ball onto the real axis. 
Then its (quasi-)stability can be understood by the approximate conservation of the number of particles in the NREFT. 
If the gradient energy hits the pole of the propagator of the relativistic field, 
it gives imaginary terms in the Lagrangian of the EFT. 
This leads to the emission of relativistic particles and hence the decay of the oscillon. 
We have checked that the lifetime is exponentially suppressed 
with a large exponent which is inversely proportional to a small expansion parameter.

We have also discussed stability against small perturbations
and see that the Q-ball analogy works well.
As a result, we have found even with gravitational interaction,
a necessary and sufficient condition 
for the stability
against small perturbations is simply given
by $\dd Q/\dd \omega<0$.

As an example, 
we have considered an axion-like potential and found oscillon solutions 
taking into account the gravity effect. 
Our results are consistent with the ones in the literature, but 
can be extended to the regime where relativistic corrections become relevant. 
We have found that 
a long-lived oscillon solution is absent in the limit where the gravity effect vanishes. 
The lifetime of axion stars can be estimated from these results 
and we have found that it is much shorter than the present age of the universe 
for the dense axion star 
but is much longer for the dilute axion star (in accordance with the previous results of \cite{Eby:2015hyx}). 
We have concluded that dilute axion stars survive until the present day for realistic parameters of axion.

We also discuss the consistency of our EFT with the ones proposed by BMZ and NGK in Refs.~\cite{Braaten:2016kzc, Namjoo:2017nia, Braaten:2018lmj}. As they have shown in those papers, we have checked that all of the EFTs are consistent, though depending on the application, different approaches have different advantages. In particular, one of the relatively low-frequency modes is integrated out in the EFTs of BMZ and NGK, but is not integrated out in our EFT. Since the low-frequency mode gives a large correction to the NREFT, our method obtains a smaller correction than that of BMZ and NGK. Even if we do not integrate out that relatively low-frequency mode, we can easily calculate the profile of scalar field configuration by solving the equation of motion numerically. 

We have pointed out that, in a system where gravity contributes at leading order in the equation of motion, it is not consistent to account for relativistic corrections without taking into account corrections to the gravitational interaction as well. This is the situation in a typical dilute boson star, where gravity is one of the dominant forces determining the bound state configuration. The EFT we  have presented here, as well as that of NGK \cite{Namjoo:2017nia} and BMZ \cite{Braaten:2018lmj}, does not calculate corrections to the gravitational interaction. Interestingly, these corrections can give rise to new interactions and potentially new decay diagrams mediated by gravity and self-interactions. We leave a full analysis of this topic for a future work.

\section*{Acknowledgements}

MY is grateful to A. H. Guth and M. H. Namjoo for stimulating discussions. The authors also thank K. Blum, H. Kim, and P. Suranyi for useful comments. The work of JE was supported by the Zuckerman STEM Leadership Program.

\begin{appendix}

\section{Corrections for $\phi^6$ theory}
\label{app:phi6}

Here, we show corrections up to $\mathcal{O}(\epsilon^3)$ for a $\phi^6$ theory, 
where the relativistic potential is given by 
\beq
 V_{\rm int}(\phi^2) = \frac{\lambda_4}{4 !} \phi^4 + \frac{\lambda_6}{6! m_\phi^2} \phi^6. 
 \label{V2}
\eeq
The effective potential is 
\beq 
 \left. V_{\rm eff} (\abs{\psi_1})\right\vert_{\epsilon_V^2} = 
 \frac{\lambda_4}{(2!)^2} \abs{\psi_1}^4 
 + \frac{\lambda_6 + 1/8 \lambda_4^2}{(3!)^2m_\phi^2} \abs{\psi_1}^6 
 + \frac{\lambda_4 \lambda_6}{(4!)^2m_\phi^4} \abs{\psi_1}^{8} 
 + \frac{19}{6}\frac{\lambda_6^2}{(5!)^2m_\phi^6} \abs{\psi_1}^{10},
 \label{eff2}
\eeq
up to $\epsilon_V^2$ 
and 
\beq
  \left. V_{\rm eff} (\abs{\psi_1})\right\vert_{\epsilon_V^3} = 
  \frac{1}{4} \frac{\lambda_4^3}{(4 !)^2 m_\phi^4} \abs{\psi_1}^8 
  + \frac{55}{24} \frac{\lambda_4^2 \lambda_6}{(5 !)^2 m_\phi^6} \abs{\psi_1}^{10}
  + \frac{241}{16} \frac{\lambda_4 \lambda_6^2}{(6 !)^2 m_\phi^8} \abs{\psi_1}^{12}
  + \frac{6027}{32} \frac{\lambda_6^3}{(7 !)^2 m_\phi^{10}} \abs{\psi_1}^{14}, 
\eeq
up to $\epsilon_V^3$. 

The corrections involving $\epsilon_t$ or $\epsilon_x$ start from $\epsilon_V^2 \epsilon_t$ or $\epsilon_V^2 \epsilon_t$: 
\beq
 &&\left. V_{\rm eff}  \right\vert_{\epsilon_V^2 \epsilon_t, \epsilon_V^2 \epsilon_x} 
 = 
 \frac{1}{8} \frac{\lambda_4^2}{(3!)^2 m_\phi^2} \psi_1^{\dagger3} D_3 \psi_1^3 
 + \frac{1}{2} \frac{\lambda_4 \lambda_6}{(4!)^2 m_\phi^4} \lmk \psi_1^{\dagger3} D_3 (\psi_1^\dagger \psi_1^4) + \psi_1^{\dagger4} \psi_1 D_3 \psi_1^3 \rmk 
 \nn
 &&~~~~~~~~~~~~~~~~~~~~~~~~~~~~~~~~~~~~+ \frac{1}{24} \frac{\lambda_6^2}{(5!)^2 m_\phi^6} \lmk \psi_1^{\dagger5} D_5 \psi_1^5  
 + 75 \psi_1^{\dagger4} \psi_1 D_3 (\psi_1^\dagger \psi_1^4) \rmk, 
\eeq
where $D_n$ is defined by \eq{Dn}.

\section{Construction up to $\epsilon^3$ order} \label{app:eps3}
Here, as an illustration, we show how to construct the effective Lagrangian by adopting a concrete example
where the potential is given by $\lambda_4\phi^4/4!$ 
and there is no gravitational interaction.
We assume $\delta_t\sim\delta_x\sim\delta_V$ and
will construct the effective Lagrangian up to $\epsilon^3$ order.

\paragraph{Diagrammatic derivation.}
First we list all the interaction vertices after the decomposition given in Eq.~\eqref{decomposition1}. The interactions between the NR mode and relativistic mode are given by
\begin{align}
\mbox{
	\includegraphics[width=0.85\textwidth]{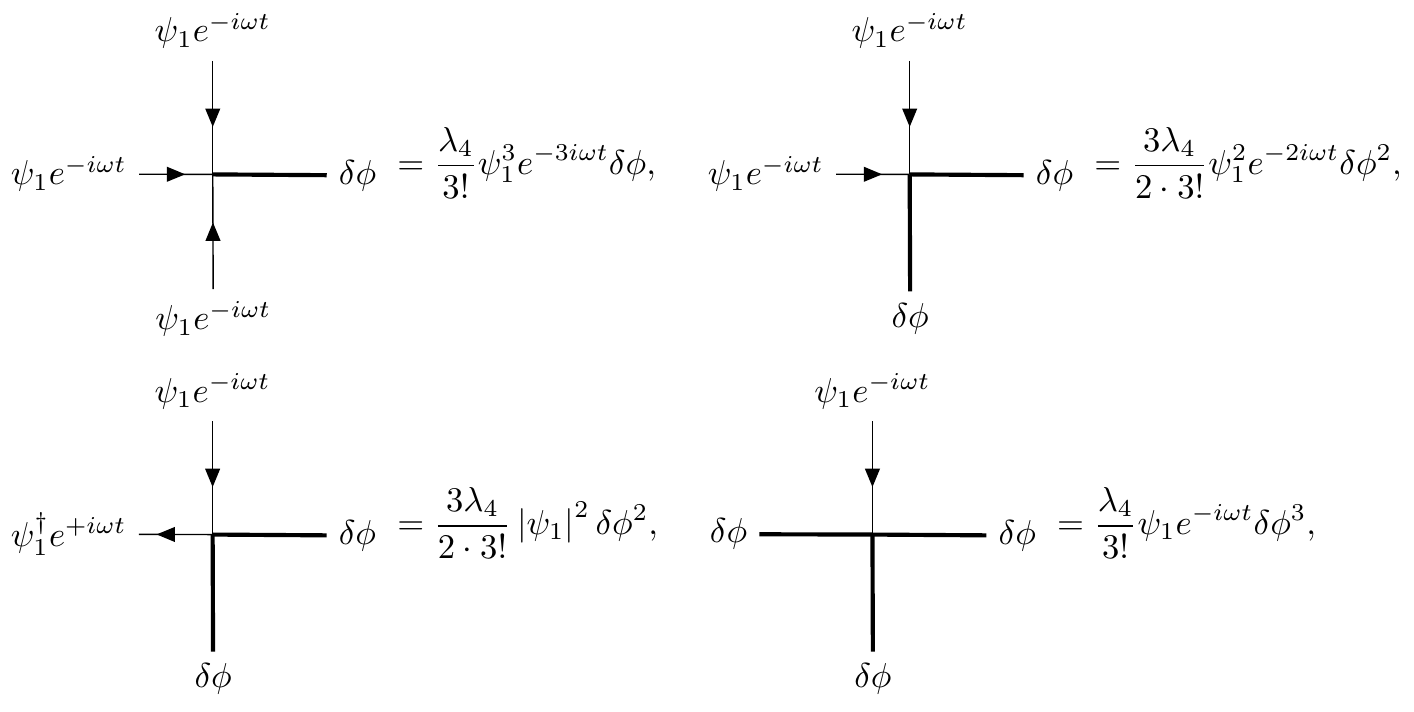}
}
\end{align}
and their conjugates. There also exists a self interaction for the relativistic mode,
\begin{align}
\mbox{
	\includegraphics[width=0.3\textwidth]{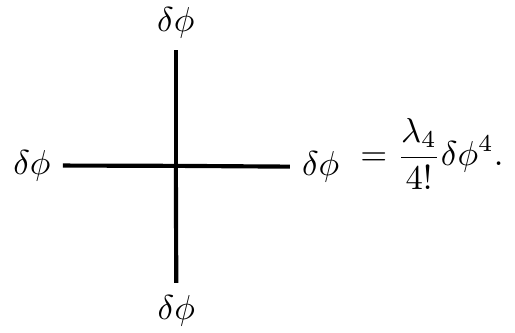}
}
\end{align}
See also Eqs.~\eqref{nr_rela} and \eqref{rela_self}.

Then, to get the NREFT, all one has to do is to integrate out the relativistic mode $\delta \phi$. Throughout this paper, we focus on the \textit{classical} NREFT, and hence all the diagrams that will be integrated out must be \textit{tree} diagrams.
The leading order term in the coupling expansion is obtained from
\begin{align}
\mbox{
	\includegraphics[width=0.99\textwidth]{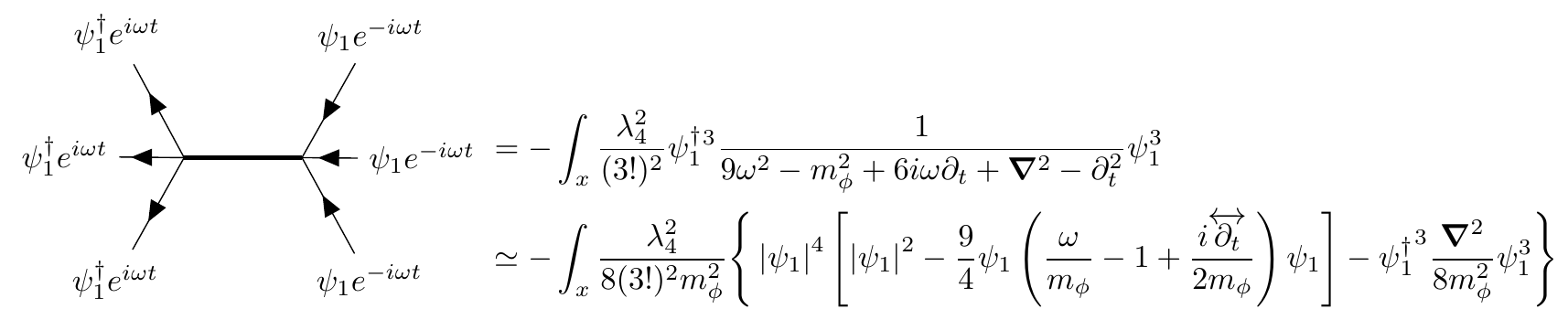}
}
\end{align}
where we dropped terms of higher order than $\epsilon^3$ and also the imaginary part in the second line.
The next to leading order term involves three vertices as depicted below
\begin{align}
\mbox{
	\includegraphics[width=0.6\textwidth]{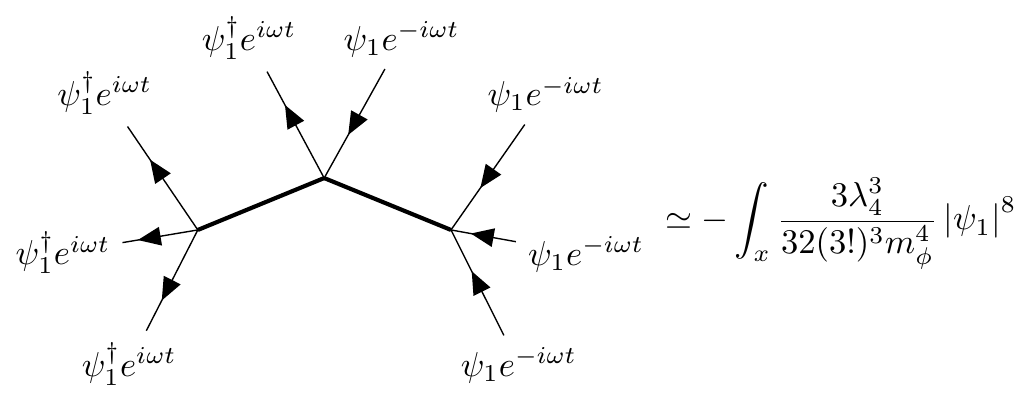}
	}
\end{align}
where again we dropped terms of higher order than $\epsilon^3$ and also the imaginary part.
To sum up, the effective potential up to $\epsilon^3$ is given by
\begin{align}
	V_\text{eff}^{(\epsilon^3)} = &
	\frac{\lambda_4}{4} \abs{\psi_1}^4 
	+ \frac{\lambda_4^2}{8(3 !)^2} \abs{\psi_1}^6
	+ \frac{3 \lambda_4^3}{32 (3 !)^3 m_\phi^4} \abs{\psi_1}^8 \nonumber \\
	&- \frac{\lambda_4^2}{8 (3 !)^2 m_\phi^2} \left[ \frac{9}{4} \abs{\psi_1}^4 \psi_1 \left( \frac{\omega}{m_\phi} - 1+ \frac{i \overleftrightarrow{\partial_t}}{2 m_\phi} \right) \psi_1
	+ \psi{_1^\dag}^3 \frac{\bm{\nabla}^2}{8 m_\phi^2} \psi_1^3
	\right].
\end{align}

\paragraph{Derivation from EoM.}
Here we present another way to derive the NREFT which is essentially equivalent but still useful.
Starting from the equation of motion (EoM), we solve $\delta \phi$ order by order in terms of $\epsilon_V$.
In this case,
the original EOM is given by
\begin{align}
\label{app:edu:eom}
         \left(\Box+m_\phi^2\right)\phi+\frac{\lambda_4}{3!}\phi^3=0.
\end{align}
We decompose $\phi$ as 
\begin{align}
         \phi(t, \bm{x})= 
         \sum_{n>0} e^{-in \omega t}\psi_n(t, \bm{x})+\text{h.c.}, 
\label{app:decomposition1}
\end{align}
and assume $\psi_1$ dominates $\phi$.
Then, at $\epsilon_V$ order, we have
\begin{align}
         \left(\Box+m^2_\phi\right)\delta \phi^{(\epsilon_V^1)}
         =-\frac{\epsilon_V}{3!}\lambda_4 \psi^3_1 e^{-3i\omega t}+{\rm h.c.}, 
\end{align}
and the resulting solution is
\begin{align}
         \psi_3^{(\epsilon_V^1)}=\frac{\epsilon_V\lambda_4}{3!\cdot8m^2_\phi}
         \left(1-\epsilon_t\frac{3(3(\omega-m_\phi)+i\partial_t)}{4m_\phi}
         -\frac{\epsilon_x{\bm\nabla}^2}{8m^2_\phi}
         \right)\psi_1^3+\mathcal{O}{(\epsilon^3)}
         .
\end{align}
By substituting $\phi=[e^{-i\omega t}\psi_1+e^{-3i\omega t}\psi_3^{(\epsilon_V^1)}+\text{h.c.}]$ into EOM,
we obtain the EOM at $\epsilon_V^2$ order.
In this order,
only $\psi_3^{(\epsilon_V^2)}$ contributes to effective action at $\mathcal{O}(\epsilon^3)$.
We have
\begin{align}
        \psi_3^{(\epsilon_V^2)}&=\frac{\epsilon_V\lambda_4}{8m^2_\phi}{|\psi_1|^2}\psi_3^{(\epsilon_V^1)}
        \nonumber \\
        &=\frac{\epsilon_V^2\lambda_4^2}{3!\cdot8^2m^4_\phi}|\psi_1|^2\psi_1^3+\mathcal{O}(\epsilon^3). 
\end{align}
Then, by substituting $\phi=[e^{-i\omega t}\psi_1+e^{-3i\omega t}(\psi_3^{(\epsilon_V)}+\psi_3^{(\epsilon_V^2)})+\text{h.c.}]$ into EOM,
we obtain the effective EOM for $\psi\equiv e^{-i\omega t}\psi_1$
\begin{align}
         \left(\Box+m^2_\phi\right)\psi=
         &-\frac{\epsilon_V\lambda_4}{2}|\psi|^2\psi-\frac{\epsilon_V^2\lambda_4^2}{96m^2_\phi}|\psi|^4\psi
        -\frac{\epsilon_V^3\lambda_4^3}{576m^4_\phi}|\psi|^6\psi\nonumber \\
        &+\frac{\epsilon_V^2\lambda_4^2}{96m^2_\phi}
        \left[
        \frac{9\epsilon_t}{4m_\phi}|\psi|^4(i\partial_t-m_\phi)\psi
        +{\psi^\dagger}^2\frac{\epsilon_x{\bm\nabla}^2}{8m^2_\phi}\psi^3
        \right]. 
\end{align}
Then, we can obtain the effective action at $\mathcal{O}(\epsilon^3)$ order:
\begin{align}
         \mathcal{L}_{\rm eff}
         &=|\partial \psi|^2-m^2_\phi|\psi|^2-V_{\rm eff}^{(\epsilon^3)},\\
         V_{\rm eff}^{(\epsilon^3)}
         &=
         \frac{\lambda_4}{4}|\psi|^4
         +\frac{\lambda_4^2}{288m^2_\phi}|\psi|^6
        +\frac{\lambda_4^3}{2304m^4_\phi}|\psi|^8\nonumber \\
        &-\frac{\lambda^2_4}{96m^2_\phi}
        \left[
        \frac{3}{4m_\phi}|\psi|^4\psi^\dagger(i\partial_t-m_\phi)\psi
        +{\psi^\dagger}^3\frac{{\bm\nabla}^2}{24m^2_\phi}\psi^3
        \right].
\end{align}

\section{The stability condition against small perturbations}
\label{app:stab}
Here, we derive the  stability condition against small perturbations
when gravitational interactions are included.
We fundamentally follow stability discussions done in~\cite{Lee:1991ax,Tsumagari:2009zp}.
The Lagrangian of the system can be written as
\footnote{
We restrict ourselves to the effective theory at leading order in $\epsilon_{g,x,t}$
because in that case the derivative parts in the effective action remain in canonical forms.
}
\begin{align}
         \mathcal{L}&=|\dot{\psi}|^2-\omega^2|\psi|^2-S_3(\omega)\\
         &\equiv K-S_3(\omega),\\
         S_3(\omega) &\equiv \int_{\bm x}
         \left|{\bm\nabla}\psi\right|^2
    +   \left(m^2_\phi (1-2\Psi)-\omega^2\right)
         |\psi|^2+V_{\rm eff}(|\psi|^2)
	+\frac{\left({\bm\nabla}\Psi\right)^2}{8\pi G}, 
\end{align}
where $K$ denotes a kinetic term
and $\Psi$ is a gravitational potential.
The dot denotes derivative with respect to time.
Suppose that the Q-ball solution is given by $e^{-i\omega t}\sigma(r)/\sqrt{2}$ where $\sigma(r)$
is a real function.
$S_3(\omega)$ is stationarized by the Q-ball solution $\sigma(r)/\sqrt{2}$.
In second order in fluctuations, we can find eigenvectors and eigenvalues:
\begin{align}
         S_3''\begin{pmatrix} \delta \psi_a\\\delta\Psi_a\end{pmatrix}
         =\lambda_a\begin{pmatrix} \delta \psi_a\\\delta\Psi_a\end{pmatrix}.
\end{align}
We normalize them as follows
\begin{align}
         \int_{\bm x} \begin{pmatrix}  \delta \psi_a^\dagger &\delta\Psi_a^\dagger\end{pmatrix}
         \begin{pmatrix} \delta \psi_b\\\delta\Psi_b\end{pmatrix}=\delta_{ab}. 
\end{align}
Note that real and imaginary parts of $\delta\psi$ are separated.
We decompose functional space into real and imaginary directions as $\delta\psi=\delta \psi_R+i\delta \psi_I$.
We know that $S_3(\omega)$ has one negative mode in real direction.
Below, we assume $S_3(\omega)$ has only one negative mode, which
is ensured if the Q-ball solution is a solution of the equation of motion with the minimal 
action $S_3$.
In addition,
there are four zero modes for $S_3''$: three of them are
related to spatial shift which we denote $\delta \psi_{S,k}~(k=1,2,3)$ and one of them corresponds to a $U(1)$ shift which we denote $\delta\psi_{\theta}$.
$\delta \psi_{S,k}$ are a real function and $\delta\psi_\theta$ is a purely 
imaginary function.
We remove such zero modes from $\delta \psi$ space and treat them in a different way.

Since the real direction has one negative mode,
the stability is non trivial.
We concentrate on eigenvectors with a real $\delta\psi_a$.
As we will see, the negative mode can be shifted to a positive mode due to the mixing with the zero mode.
We consider fluctuations in the real direction and a zero mode%
\footnote{$\delta \psi_{S,k}$ do not contribute to this analysis (see~\cite{Lee:1991ax,Tsumagari:2009zp}).}:
\begin{align}
         \psi=\frac{1}{\sqrt{2}}e^{-i\omega t -i\theta(t)}\left(\sigma (r)+\delta\psi_R({\bm x})\right),
\end{align}
where $\theta{(t)}$ denotes some real function,
which is a zero mode in $S_3$.

What is non trivial here is the mixing between the zero mode $\theta(t)$ and
the real direction in $K$.
We can solve the equation of motion for $\dot{\theta}$ and the resulting kinetic
term becomes
\begin{align}
         K&=\frac{1}{2}\dot{\delta \psi_R}^2-\frac{2\omega^2}{I}\left(
         \int_{{\bm x}} \delta \psi_R \sigma
         \right)^2,\\
         I&\equiv \int_{{\bm x}} \sigma^2.
\end{align}
We expand the vector $(\sigma (r),0)^t$ in terms of 
eigenvectors:
\begin{align}
          \begin{pmatrix} \sigma(r) \\ 0 \end{pmatrix}
          =\sum_a \sigma_a 
          \begin{pmatrix}
           \delta \psi_a \\ \delta\Psi_a\end{pmatrix}.
\end{align}
Here $\delta \psi_a$ are real functions in eigenmodes.
We define matrix $B$ and $C$ as
\begin{align}
         B_{ab}\equiv \int_{{\bm x}} \delta \psi_a\delta \psi_b,\\
         C_{ab}\equiv \int_{{\bm x}} \delta\Psi_a\delta \Psi_b,\\
         B_{ab}+C_{ab}=\hat{I}_{ab}.
\end{align}
For $(\sigma(r),0)^t$, we expect
\begin{align}
\label{ap:eq:b}
         \sigma_a=B_{ab}\sigma_b.
\end{align}

We expand $\delta \psi$ as follows
\begin{align}
          \begin{pmatrix} \delta \psi_R \\\delta\Psi \end{pmatrix}
          =\sum_a c_a(t)\begin{pmatrix} \delta \psi_a \\\delta\Psi_a \end{pmatrix}.
\end{align}
For $c_a(t)$, the Lagrangian can be written as
\begin{align}
         \mathcal{L}&=\frac{1}{2}\dot{c}(t)^tB\dot{c}(t) - \frac{2\omega^2}{I}V^2-\sum_a
         \frac{ \lambda_a c_a(t)^2}{2},\\
        V &\equiv \int_{{\bm x}} \delta \psi_R \sigma=\sum_{a,b}\sigma_aB_{ab}c_b(t)=\sum_a\sigma_ac_a(t),\\
\end{align}
Since $B_{ab}$ is a positive matrix,
the minimal eigenvalue of the following matrix
\begin{align}
         M_{ab}\equiv \lambda_a\delta_{ab}+\frac{4\omega^2}{I}\sigma^t\sigma,
\end{align}
determines the stability.
If the minimal eigenvalue is positive, we expect that Q-ball configuration is 
stable against small perturbations. 
On the contrary, if the minimal eigenvalue is negative, Q-ball configuration
becomes unstable agains small perturbations.
We denote the eigenvalues of $M$ by $\{\Lambda_a\}$,
and require that
$(\Lambda_1<\Lambda_2<...)$ and also $(\lambda_1<\lambda_2<...)$
are different.
\footnote{One may worry about the degeneracy of eigenmodes. However, if we deform the potential infinitesimally, such a degeneracy is broken in general. The physical results are expected to be unaffected by such an infinitesimal deformation.}
Then, we have
\begin{align}
         \det |M-z\hat{I}|\propto G(z)\equiv 1+\frac{4\omega^2}{I}\sum_a \frac{\sigma_a\sigma_a}{\lambda_a-z}.
\end{align}
Here, we use the relation
\begin{align}
        \det|\hat{I}+uu^t|=1+u^tu,
\end{align}
with $u$ being some vector.
Note that
\begin{align}
         \frac{{\rm d} G}{{\rm d} z}>0,~~\lim_{z\rightarrow \lambda_j\pm 0}G(z)=\mp \infty.
\end{align}
These relations ensure that there is only one $\Lambda_a$ between $\lambda_a$ and $\lambda_{a+1}$.
Note that
$\lambda_a$ has only one negative mode which we denote $\lambda_1$.
The condition $\Lambda_1>0$ is now equivalent to
\begin{align}
         G(0)<0.
\end{align}

As is in~\cite{Lee:1991ax,Tsumagari:2009zp}, we can connect
$G(0)$ and $dQ/d\omega$ and show that the stability condition is given by
\begin{align} \label{stabilityApp}
         \frac{\omega}{Q}\frac{{\rm d} Q}{{\rm d} \omega}<0.
\end{align}
Below, we derive $G(0)= \frac{\omega}{Q}\frac{{\rm d} Q}{{\rm d} \omega}$.

First, we note that $\sigma(\omega)$ is a solution to the equation of motion derived by $S_3(\omega)$. Taking the derivative of the equations of motion for $\psi$ and $\Psi$ with respect to $\omega$, we have 
\begin{align}
\label{ap:eq:f1}
         S_3'' \begin{pmatrix}
         \frac{\partial\sigma}{\partial\omega}\\
         \frac{\partial \Psi}{\partial\omega}
         \end{pmatrix}=
         \begin{pmatrix}
         2\omega\sigma\\ 0
         \end{pmatrix}.
\end{align}
Multiplying $\sum_a\sigma_a$ on 
$\frac{1}{\lambda_a}S_3''(\delta{\psi}_a~\delta\Psi_a)^t=(\delta{\psi}_a~\delta\Psi_a)^t$,
we have
\begin{align}
\label{ap:eq:f2}
         S_3''\sum_a\frac{\sigma_a}{\lambda_a}\begin{pmatrix}
         \delta\psi_a \\\delta\Psi_a
         \end{pmatrix}
         =\begin{pmatrix}
         \sigma\\0
         \end{pmatrix}.
\end{align}
Since $\delta\psi_a$ and $\del \sigma / \del \omega$ do not contain the zero mode of $S_3$, we can multiply the inverse matrix of $S_3$ in (\ref{ap:eq:f1}) and (\ref{ap:eq:f2}). Then, we have

\begin{align}
         2\omega\sum_a \frac{\sigma_a}{\lambda_a}
         \delta\psi_a =\frac{\partial \sigma}{\partial \omega}.
\end{align}
Multiplying $\int_{\bm x}\sum_b\sigma_b\delta\psi_b$, we obtain
\begin{align}
         2\omega \sum_a \frac{\sigma_a\sigma_a}{\lambda_a}=\int_{\bm x} \sigma
         \frac{\partial \sigma}{\partial \omega},
\end{align}
where we use (\ref{ap:eq:b}).
 On the other hand, by differentiating $Q=\omega\int_{{\bm x}}~\sigma^2$ with respect to $\omega$, we have
\begin{align}
         \frac{\omega}{Q}\frac{\partial Q}{\partial \omega}=1+\frac{2\omega}{I}\int \sigma
         \frac{\partial \sigma}{\partial \omega}.
\end{align}
Taken together, we have
\begin{align}
         \frac{\omega}{Q}\frac{\partial Q}{\partial \omega}
         =1+\frac{4\omega^2}{I}\sum_a \frac{\sigma_a\sigma_a}{\lambda_a}=G(0).
\end{align}
Therefore, we conclude that stability is ensured when Eq. \eqref{stabilityApp} is satisfied.

\end{appendix}

\bibliography{reference}{}

\end{document}